\documentstyle[epsfig]{mn}

\newif\ifAMStwofonts
\AMStwofontstrue

\ifoldfss
  \newcommand{\rmn}[1] {{\rm #1}}

  \ifCUPmtlplainloaded \else
    \NewTextAlphabet{textbfit} {cmbxti10} {}
    \NewTextAlphabet{textbfss} {cmssbx10} {}
    \NewMathAlphabet{mathbfit} {cmbxti10} {} 
    \NewMathAlphabet{mathbfss} {cmssbx10} {} 
  \fi
  \ifAMStwofonts
    \ifCUPmtlplainloaded \else
      \NewSymbolFont{upmath} {eurm10}
      \NewSymbolFont{AMSa} {msam10}
      \NewMathSymbol{\upi}     {0}{upmath}{19}
      \NewMathSymbol{\umu}     {0}{upmath}{16}
      \NewMathSymbol{\upartial}{0}{upmath}{40}
      \NewMathSymbol{\leqslant}{3}{AMSa}{36}
      \NewMathSymbol{\geqslant}{3}{AMSa}{3E}

      \let\leq=\leqslant 
      \let\geq=\geqslant \let\ge=\geqslant
    \fi
  \fi
\fi 

\ifnfssone
  \newmathalphabet{\mathit}
  \addtoversion{normal}{\mathit}{cmr}{m}{it}
  \addtoversion{bold}{\mathit}{cmr}{bx}{it}
  \newcommand{\rmn}[1] {\mathrm{#1}}

  \newmathalphabet{\mathbfit} 
  \addtoversion{normal}{\mathbfit}{cmr}{bx}{it}
  \addtoversion{bold}{\mathbfit}{cmr}{bx}{it}
  \newmathalphabet{\mathbfss} 
  \addtoversion{normal}{\mathbfss}{cmss}{bx}{n}
  \addtoversion{bold}{\mathbfss}{cmss}{bx}{n}
  \ifAMStwofonts
    \ifCUPmtlplainloaded \else
      \UseAMStwoboldmath
      \makeatletter
      \new@mathgroup\upmath@group
      \define@mathgroup\mv@normal\upmath@group{eur}{m}{n}
      \define@mathgroup\mv@bold\upmath@group{eur}{b}{n}
      \edef\UPM{\hexnumber\upmath@group}
      \new@mathgroup\amsa@group
      \define@mathgroup\mv@normal\amsa@group{msa}{m}{n}
      \define@mathgroup\mv@bold\amsa@group{msa}{m}{n}
      \edef\AMSa{\hexnumber\amsa@group}
      \makeatother
      \mathchardef\upi="0\UPM19
      \mathchardef\umu="0\UPM16
      \mathchardef\upartial="0\UPM40
      \mathchardef\leqslant="3\AMSa36
      \mathchardef\geqslant="3\AMSa3E

      \let\leq=\leqslant 
      \let\geq=\geqslant \let\ge=\geqslant
    \fi
  \fi
\fi 

\ifnfsstwo
  \newcommand{\rmn}[1] {\mathrm{#1}}

  \DeclareMathAlphabet{\mathbfit}{OT1}{cmr}{bx}{it}
  \SetMathAlphabet\mathbfit{bold}{OT1}{cmr}{bx}{it}
  \DeclareMathAlphabet{\mathbfss}{OT1}{cmss}{bx}{n}
  \SetMathAlphabet\mathbfss{bold}{OT1}{cmss}{bx}{n}
  \ifAMStwofonts
    \ifCUPmtlplainloaded \else
      \DeclareSymbolFont{UPM}{U}{eur}{m}{n}
      \SetSymbolFont{UPM}{bold}{U}{eur}{b}{n}
      \DeclareSymbolFont{AMSa}{U}{msa}{m}{n}
      \DeclareMathSymbol{\upi}{0}{UPM}{"19}
      \DeclareMathSymbol{\umu}{0}{UPM}{"16}
      \DeclareMathSymbol{\upartial}{0}{UPM}{"40}
      \DeclareMathSymbol{\leqslant}{3}{AMSa}{"36}
      \DeclareMathSymbol{\geqslant}{3}{AMSa}{"3E}

      \let\leq=\leqslant 
      \let\geq=\geqslant \let\ge=\geqslant
    \fi
  \fi
\fi 

\ifCUPmtlplainloaded \else
  \ifAMStwofonts \else 
    \def\upi{\pi}
    \def\umu{\mu}
    \def\upartial{\partial}
  \fi
\fi

\newcommand{\ud}{{\rmn{d}}}
\newcommand{\be}{{\bmath{e}}}
\newcommand{\clp}{{\mathcal{P}}}
\newcommand{\clb}{{\mathcal{B}}}
\newcommand{\cln}{{\mathcal{N}}}

\title[Harmonic analysis of CMB data]{Harmonic analysis of cosmic microwave
background data II: from ring-sets to the sky}
\author[A.\ D.\ Challinor et al.]{%
Anthony D.\ Challinor,$^1$\thanks{E-mail: a.d.challinor@mrao.cam.ac.uk}
Daniel J.\ Mortlock,$^{1,2}$ Floor van Leeuwen,$^{2}$
\newauthor
Anthony N.\ Lasenby,$^1$ Michael P.\ Hobson,$^1$ Mark A.\ J.\ Ashdown$^{1,2}$
\newauthor
and George P.\ Efstathiou$^2$
\\
$^1$Astrophysics Group, Cavendish Laboratory, Madingley Road,
Cambridge CB3 0HE, U.K. \\
$^2$Institute of Astronomy, Madingley Road, Cambridge CB3 0HA, U.K.}

\date{Accepted by MNRAS: 11 December 2001}

\pagerange{\pageref{firstpage}--\pageref{lastpage}}
\pubyear{2001}

\label{firstpage}

\begin{document}

\maketitle

\begin{abstract}
Despite the fact that the physics of the cosmic microwave background
anisotropies is most naturally expressed in Fourier space, pixelised maps
are almost always used in the analysis and simulation of microwave data.
A complementary approach is investigated here, in which maps are used only
in the visualisation of the data, and the temperature anisotropies and
polarization are only ever expressed in terms of their spherical multipoles.
This approach has a number of advantages: there is no information loss
(assuming a band-limited observation); deconvolution of asymmetric beam
profiles and the temporal response of the instrument are naturally included;
correlated noise can easily be taken into account, removing the
need for additional `destriping'; polarization is also analysed in the same
framework; and reliable estimates of the spherical multipoles of the sky and
their errors are obtained
directly for subsequent component separation and power spectrum estimation.
The formalism required to
analyse experiments which survey the full sky by scanning on circles
is derived here, with particular emphasis on the \emph{Planck} mission.
A number of analytical results are obtained in the limit of simple scanning
strategies.
Although there are non-trivial computational obstacles to be overcome before
the techniques described here can be implemented at high resolution, if these
can be overcome the method should allow for a more robust return from the next
generation of full-sky microwave background experiments.

\end{abstract}


\section{Introduction}
\label{sec:intro}

The cosmic microwave background (CMB) represents the single most powerful probe
of the early universe available to modern astronomy. The pioneering results of
the \emph{Cosmic Background Explorer} (\emph{COBE}; Smoot et al.\ 1992)
are being extended by the current generation of ground-based and balloon-borne
experiments (e.g.\ Scott et al.\ 1996; Tanaka et al.\ 1996;
Netterfield et al.\ 1997; de Oliveira-Costa et al.\ 1998; Coble et al.\ 1999;
de Bernardis et al.\ 2000; Hanany et al.\ 2000; Wilson et al.\ 2000;
Padin et al.\ 2001), but it is the upcoming
satellite experiments that are set to revolutionise the
field. By the end of 2002 the
\emph{Microwave Anisotropy Probe}\footnote{http://map.gsfc.nasa.gov/}
(\emph{MAP}) will have observed the entire sky with a
resolution of $\ge 0\fdg 21$
and \emph{Planck}\footnote{http://astro.estec.esa.nl/Planck/},
scheduled for launch in 2007, should have the
capacity to produce high sensitivity, all-sky maps to a resolution of 
$\sim$ 0\fdg 1 in four of its ten frequency channels. 
However, it is not sufficient merely to obtain such 
extraordinary measurements; careful analysis of the data-sets is 
critical if these missions are to fulfill their promised scientific goals.

Most present techniques for CMB data analysis employ pixelised
maps of the sky. Both Bond et al.~\shortcite{bond99} and
Borrill~\shortcite{borrill99} describe methods to create maximum-likelihood
maps from time-ordered data and telescope pointing information,
and these techniques have been successfully applied in the analysis
of the BOOMERanG~\cite{deb00,net01} and MAXIMA~\cite{hanany00,lee01}
observations.
From the maps and their uncertainties, the power spectrum of the 
microwave sky can be estimated in a number of different ways
(Bond, Jaffe \& Knox 1998; Borrill 1999; Oh, Spergel \& Hinshaw 1999;
Wandelt, Hivon \& G\'{o}rski 2001), again employing maximum-likelihood
techniques. These methods have a number of useful features:
a map represents vast data compression relative to the time-ordered data
(e.g.\ Borrill 1999), but is still a sufficient statistic for cosmological
parameter estimation; there are usually negligible pixel-pixel noise
correlations in the beam-smoothed map; maps provide important `reality checks',
and can be inspected by eye; and unobserved or contaminated regions of the sky
can be simply removed from subsequent analysis by excluding the associated
pixels (e.g.\ Oh et al.\ 1999). From a more practical point of view, the
conventional
map-making method is `tried and tested' now. However, the use of pixelised maps
during the data analysis also involves some compromises:
the real microwave sky does not consist of a number of regions of 
uniform temperature (the prior hypothesis used to make a map), 
and so information is lost in map-making; subsequent steps in the analysis
pipeline, such as component separation, cannot easily be performed
efficiently in
real space~\cite{hobson98}; there is no single optimal or obvious choice of
pixelisation scheme; and deconvolution of the (asymmetric) beam profile and
temporal response of instrument, and removal of scan-synchronous instrument
effects require awkward additional processing during
map-making~\cite{del98a,revenu00,wu01}.
Another potential problem that has received only limited attention
thus-far (e.g. Tegmark \& de Oliveira-Costa 2001)
is how to create maps from polarization data (i.e.\ a map for each of the
Stokes parameters, or their gradient and curl components) -- there is currently
no robust algorithm for the treatment of several polarization-specific
systematic effects that plague CMB polarimetry experiments.

It is with these points in mind that a complementary,
harmonic method of data analysis is presented here. For full-sky surveys
the data analysis process would proceed from the time-ordered data 
directly to the set of spherical multipole moments, which would take the
place of a real-space map. For band-limited observations (effectively ensured
by the finite experimental beam), this harmonic reconstruction of the sky
involves essentially no loss of information. The separation of the various
astrophysical components of the microwave sky could then be performed
quite naturally in multipole space (e.g.\ Hobson et al.\ 1998), and, if
necessary, the Galactic plane could be removed while retaining orthonormality
of the basis set (G\'{o}rski 1994; Mortlock, Challinor \& Hobson 2001;
see also Section~\ref{subsec:power}).
Power spectrum estimation could then proceed naturally 
from this point; for instance the efficient method presented by
Oh et al.~\shortcite{oh99} could be used, although some generalisation would
be required to remove the dependence on forward and inverse fast spherical
transforms, which Oh et al.\ use to apply the inverse noise matrix efficiently
in the space of multipoles and to reduce memory requirements.
The multipole moments of the polarized components could be obtained using an
almost identical formalism; obviously it would incur an additional overhead for
each component to be reconstructed, but no further conceptual
development would be required.

We have endeavoured to present the harmonic
method as an end-to-end solution for the processing of time-ordered data
into power spectra. For such an analysis pipeline the benefits of clear
error propagation from timelines to power spectra afforded by our harmonic
methods are maximised. However, we do not pretend that all aspects of CMB
analysis are best performed in the harmonic domain; instead we view harmonic
methods as being complementary to standard map-based techniques.
An obvious example where map-based processing is clearly desirable
is the analysis of localised foregrounds (or CMB features).
In addition, the inversion from observations
over only a fraction of the sky to the spherical multipoles rapidly becomes
singular as the resolution of the observations is increased. Although it may be
possible to regularise the inversion in such cases
(e.g.\ with prior information on the power spectrum), or to adopt a basis set
adapted to the observed patch, the processing we describe in this paper is
best considered in the context of full-sky observations. Even then, some
aspects of the processing, such as cutting out contaminated regions close
to the galactic plane for power spectrum estimation, require rather cumbersome
operations if we insist on working directly with the full sky spherical
multipoles rather than a map synthesised from them.

The principles of harmonic data analysis as described above are quite
general, and could be applied to any full-sky CMB observations.
However, it is particularly suited to experiments with a circular 
scanning strategy, such as the \emph{Planck} mission.
The data are obtained by a combination of rotations (i.e.\ the motion of 
the detector across the sky) and convolutions (of the beam with the 
true microwave sky, and this subsequent signal with the temporal response
of the instrument), and these operations are most simply
expressed in the harmonic basis (Delabrouille, G\'{o}rski \& Hivon 1998a;
Challinor et al.\ 2000; Wandelt \& G\'{o}rski 2001).
Hence we propose that the time-ordered data be transformed to the
Fourier basis at the  earliest opportunity:
van Leeuwen et al.~\shortcite{van01} describe how to construct
point source catalogues and calibrated ring-sets simultaneously, in which
the data around each ring is represented in Fourier space. Some aspects of
this process are necessarily mission-specific, and, for the sake of generality,
it is assumed here that it is possible to obtain calibrated ring-sets
(and their errors) in this form.
The focus of this paper is on the subsequent `map'-making:
combining the ring-sets to obtain an estimate of the spherical
multipoles for the temperature anisotropies and polarization.
In this respect, we build on earlier work by
Delabrouille et al.~\shortcite{del98b}, who analysed the
statistics of the power spectrum of the Fourier modes on a single ring, and the
relation with the power spectrum of the underlying temperature field on the
sphere.
More recently, Wandelt \& Hansen~\shortcite{wan01a} have proposed the
`ring-torus' method for estimating the temperature power spectrum from the
two-dimensional Fourier transform of data obtained on a set of rings with
centres equally spaced on a small circle. While the general philosophy of the
`ring-torus' method coincides with that advocated here, the methodology
presented here is somewhat more general. By only making use of one-dimensional
Fourier data, we are able to deal with arbitrary ring-sets and
to include instrument effects such as the inevitable variations in the
scanning properties of the telescope~\cite{van01}. Furthermore,
since we do not proceed directly to the power spectrum, we can make full use
of Fourier-based component separation algorithms (e.g.\ Hobson et al.\ 1998)
to deal carefully with foreground contamination. The cost of maintaining
this generality is the increased computational requirements of our methods
over those that exploit special symmetries of the survey.

The general formalism is described in Section~\ref{sec:hdm}, which proceeds
from a model of the instrument to the maximum-likelihood solution for the
spherical multipoles of the temperature anisotropy and polarization, and
their associated errors. The structure of the error covariance matrix,
including polarization, is studied for some simple scan strategies in
Section~\ref{sec:scan}. Making
use of some results given in Appendix B, we are able to make contact with
several well-known analytic results for idealised experiments, previously
obtained from arguments at the level of the map. In Section~\ref{sec:disc}
we discuss a number of important issues that arise during the map-making phase,
including correcting for scan-synchronous instrument effects, the
overall calibration of the instrument to external standards, and the
treatment of low frequency noise. The latter discussion includes a novel
method for dealing with uncertainties in the low frequency noise power spectrum
or insensitivity of the experiment to the monopole. Section~\ref{sec:sub}
reviews the subsequent processing of the frequency maps, including component
separation and power spectrum estimation, within the context of the harmonic
data model. Finally, we conclude in Section~\ref{sec:conc}
by reviewing the relative merits of the harmonic method, and suggest
directions for future development.

\section{The harmonic data model}
\label{sec:hdm}

The basis of the harmonic data model is the relationship
between the one-dimensional Fourier representation of the ring data and
the spherical multipole coefficients of the underlying sky. In the absence
of instrument noise and reconstruction errors in the estimation of the
Fourier ring data, the relation between the data and the sky is determined
by the scanning strategy, the point-spread function or beam of the
telescope, and the impulse response of the instrument in the time domain.
In this section, we set up a detailed, but general, model of the instrument
and scanning strategy which defines the contribution to the data coming from
the sky.
We then include instrument noise and reconstruction errors to give the
maximum-likelihood solution for the spherical multipoles of the sky, and the
covariance of their errors.

\subsection{The microwave sky}
\label{subsec:sky}

We describe the microwave sky in terms of spherical multipoles on some fixed
basis. The brightness in total intensity $I(\be; \nu)$ from the sky when
observed along direction $\be$ at frequency $\nu$ is represented as
\begin{equation}
I(\be; \nu) = \sum_{lm} a^I_{(lm)}(\nu) Y_{(lm)}(\be), 
\label{eq:1}
\end{equation}
where the sum is over integers $l\geq 0$, and $|m| \leq l$. The (partial)
polarization of the sky is described by Stokes (brightness) parameters
$Q(\be; \nu)$, $U(\be; \nu)$, and $V(\be; \nu)$. We define the $Q$ and $U$
Stokes parameters on a basis which forms a right-handed triad with the
incoming radiation direction $-\be$: In a spherical polar coordinate
system, we use the basis $\{\bsigma_\theta, -\bsigma_\phi\}$ with
$\bsigma_r = \be$. The $V$ brightness is a scalar function, so can be
expanded in multipoles as
\begin{equation}
V(\be; \nu) = \sum_{lm} a^V_{(lm)}(\nu) Y_{(lm)}(\be).
\label{eq:2}
\end{equation}
It is convenient to combine the $Q$ and $U$ brightnesses into a symmetric,
trace-free second-rank tensor:
\begin{eqnarray}
\clp^{ab}(\be; \nu) &=& \frac{1}{2} [Q(\be;\nu)(\bsigma_\theta
\otimes\bsigma_\theta -\bsigma_\phi\otimes\bsigma_\phi) \nonumber \\
&&\mbox{} - U(\be;\nu)(\bsigma_\theta\otimes\bsigma_\phi
+ \bsigma_\phi\otimes\bsigma_\theta)],
\label{eq:3}
\end{eqnarray}
which can then be expanded in the transverse, trace-free tensor
harmonics (Kamionkowski, Kosowsky \& Stebbins 1997) as
\begin{equation}
\clp_{ab}(\be; \nu) = \sum_{lm} [a^G_{(lm)}(\nu) Y_{(lm)ab}^G(\be)
+ a^C_{(lm)}(\nu) Y_{(lm)ab}^C(\be)].
\label{eq:4}
\end{equation}
Here, the sum is over $l \geq 2$, and $|m| \leq l$. The superscripts
$G$ (for gradient, often called electric) and $C$ (for curl, often called
magnetic) refer to the two types of transverse, trace-free harmonics.
All multipoles satisfy $a^{P\ast}_{(lm)}(\nu) = (-1)^m a^{P}_{(l\, -m)}(\nu)$
since the brightnesses are real and we have adopted the Condon-Shortley phase
for the spherical harmonics.

In this paper we assume that the $a^{P}_{(lm)}(\nu)$ contain all astrophysical
components. In particular, we include the contribution of unresolved
extra-Galactic radio sources (point sources). The point source catalogue is an
important deliverable product of the \emph{Planck} mission; an
algorithm for simultaneously constructing the bright point source catalogue
and calibrating the instrument from data ordered by phase on rings is
described by van Leeuwen et al.~\shortcite{van01}. Here, we assume that
the contribution of those bright point sources identifiable on individual
rings has not been removed from the data, although
we leave open the question of whether this assumption is optimal in the
context of the harmonic data model. The steps described in this paper for
reconstructing the multipoles of the sky from the Fourier modes of the ring
data are largely independent of whether the data includes
the bright point source contribution or not. Of course, if the bright
point sources are removed from the data prior to reconstructing the sky,
our solution excludes the contribution from those identified point sources.
One practical problem with leaving bright
point sources in the ring-sets is that the sky must be represented by a
larger number of multipoles to avoid biasing the estimates of the lower
multipoles. Furthermore, it is then
essential that the point sources are removed during the component separation
phase, before any cosmological analysis of the maps
(e.g.\ Vielva et al.\ 2001).
Removing the brightest point sources at the level of the ring-sets leads to a
modest increase in the complexity of the pipeline leading from time-ordered
data to calibrated ring-sets. In particular, although a given source will only
give rise to a prominent feature on a small subset of rings, the
source's contribution must be removed consistently from every ring to ensure
that the remaining data on each ring is derived from the same underlying
sky. Point sources that are too faint to be identifiable on any single ring may
still be detectable statistically when the rings are combined into the
best-fitting set of multipoles (or map). Statistical detection of faint
point sources is easily implemented during the component separation
phase~\cite{hobson99}, and does not require the same pre-processing steps
as removal of bright point sources~\cite{vielva01}.

\subsection{Ring data}
\label{subsec:ring}

\begin{figure}
\epsfig{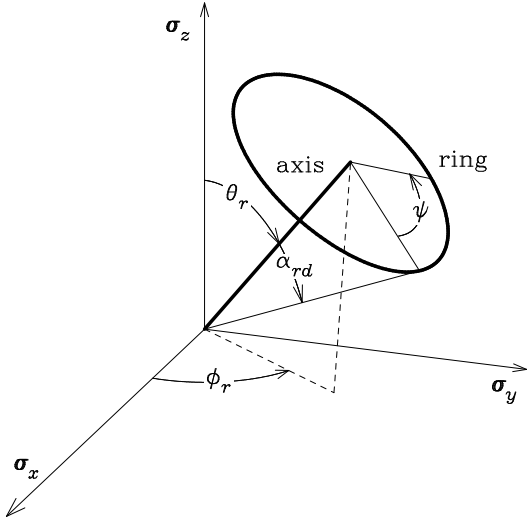}
\caption{Geometry of the $r$th ring for the $d$th detector.
The average direction of the spin axis over the $N_{\rmn s}$ revolutions that
make up the data on a single ring has polar angle $\theta_r$ and azimuth
$\phi_r$, and is the same for all detectors. The ring opening angle
$\alpha_{rd}$ depends on the detector and the ring (due to variation in the
inertia tensor during the mission). For a given ring and detector, $\psi$
measures the angle around the ring. The point $\psi=0$ is defined as the most
southerly intersection of the ring and the plane containing the ring axis and
the $z$-direction. Note that $\psi=0$ is attained at different times for
different detectors -- the phase offsets being determined by the focal plane
geometry.}
\label{fig:ring}
\end{figure}

A CMB mission with a circular scanning strategy can be described in terms of
$N_{\rmn{d}}$ sets of $N_{\rmn{r}}$ rings on the sky,
each set referring to a given detector (see Fig.~\ref{fig:ring}).
For all detectors, each ring
is covered $N_{\rmn{s}}$ times by spinning the instrument about the
axis of the ring. In the case of the \emph{Planck} mission, the satellite
will rotate $N_{\rmn{s}} \simeq 60$ times about it axis before moving on
to a new pointing. The $r$th ring is specified by its axis pointing,
$(\theta_{r},\phi_{r})$, which coincides with the average
pointing of the spin axis during the $N_{\rmn{s}}$ revolutions. The axis
pointing of a ring is the same for all detectors on the instrument. Note that
the concept of a ring only makes sense if any precession of the spin axis
(due to initial misalignment of the spin axis with the principal directions
of the inertial tensor, or external torques) can be
engineered to be insignificant compared to the smallest beam width.
The opening angle of the $r$th ring for the $d$th detector is denoted by
$\alpha_{rd}$. It is defined as the average (over
the $N_{\rmn{s}}$ revolutions) angle between the spin axis and the nominal
main beam direction on the sky of the appropriate horn in the focal plane. Not
only does $\alpha_{rd}$ depend on the detector, but it also differs between
rings for a given detector.
The dependence on the detector is determined by the focal plane geometry.
For low frequency, polarization-sensitive instruments (e.g. HEMT receivers)
there will typically be
two detectors (measuring nearly orthogonal polarization states) on a given
horn. In such an arrangement, the ring opening angle will be the same for the
two detectors, but will differ between horns. Similar comments apply to the
current state of the art designs for high frequency bolometer instruments,
where two orthogonal polarization sensitive bolometers are placed on a single
horn. The dependence of $\alpha_{rd}$ within a given detector's ring-set
arises from slow drifts of the spin axis relative to the instrument optics
as consumables are depleted during the mission (thus changing the inertia
tensor of the instrument).

The position around a given ring is specified by $\psi$, and is measured so
that $\psi=0$ is the most southerly intersection of the ring and the great
circle contaning the $z$-direction and the spin axis [see
Fig.~\ref{fig:ring}; for $\theta_r=0$ ($\pi$) we take $\psi=0$ to lie in
the $x$-$z$ plane with positive (negative) $x$].
The times at which different detectors pass through
$\psi=0$ is dependent on the focal plane geometry. For each horn in the focal
plane we define a constant reference configuration where the nominal
main beam direction is aligned with the $\bsigma_z$ direction of the
Cartesian frame used to define the spherical multipoles for the sky, and the
$z$-axis of the instrument reference system~\cite{van01} lies in the plane
normal to $\bsigma_y$, with a negative projection onto
$\bsigma_x$. (The instrument reference system, which is fixed relative to the
instrument optics, can be chosen so that its $z$-axis almost coincides with
the nutation-averaged spin axis. Variations in the inertia tensor during the
mission prevent a constant alignment of the spin and $z$-axes.) The
instrument configuration when the $d$th detector takes data at angle
$\psi$ on the $r$th ring is obtained by rotating the instrument from the
appropriate horn reference configuration. The appropriate rotation is given
by the composition\footnote{Our convention for the Euler angles
$\alpha$, $\beta$ and $\gamma$ are such that the rotation
$D(\alpha,\beta,\gamma)$ actively rotates by $\gamma$ about $\bsigma_z$,
followed by $\beta$ about $\bsigma_y$, and finally by $\alpha$ about
$\bsigma_z$ again. All rotations are right-handed. See Brink \&
Satchler~\shortcite{brink93}, whose conventions we follow, for a discussion
of the several alternatives that appear in the literature.}
$D(\phi_r,\theta_r,\psi) D(0,\alpha_{rd},\kappa_{rd})$. The
rotation $D(0,\alpha_{rd},\kappa_{rd})$ accounts for focal plane rotation
(which arises from misalignment of the spin axis and the $z$-axis of the
instrument) by: (i) first rotating the spin axis into the $x$-$z$ plane of the
sky coordinate system while leaving the main beam direction along the
$z$-direction on the sky; and (ii) taking the spin axis onto the $z$-direction
on the sky by rotating in the plane contaning the spin axis and the main beam
direction.
The angle $\kappa_{rd}$ thus measures the angle between two planes,
both containing the main beam direction, with one including the spin axis and
the other including the $z$-axis of the instrument reference system. The
rotation $D(\phi_r,\theta_r,\psi)$ takes the spin axis onto the axis
of the $r$th ring, and the main beam to the position $\psi$ around the ring.
Note that the horn reference configuration is defined by the instrument
reference system rather then the spin axis. This ensures that the beam patterns
in the given horn reference configuration are independent of variations
in the inertia tensor.

The contribution of the (time-independent) sky signal to the time-ordered data
from a given detector will be periodic in $\psi$, so the data can be co-added
in bins of $\psi$ with essentially no loss of useful information. Producing
this phase-ordered data from the time streams is a non-trivial task, since
typically the scan velocity drifts during the $N_{\rmn{s}}$ scans of
the circle: van Leeuwen et al.~\shortcite{van01} detail one possible scheme
for reconstructing the phase-ordered data and associated errors.
From the phase-ordered data $t_{rd}(\psi)$, one can estimate the Fourier
coefficients in the expansion
\begin{equation}
t_{rd}(\psi) = \sum_{n} t_{(rdn)} {\rmn{e}}^{i n \psi},
\label{eq:5}
\end{equation}
and the covariance of their errors $N_{(rdn)(r'd'n')} \equiv
\langle \Delta t_{(rdn)} \Delta t^\ast_{(r' d' n')} \rangle$.
Here, angle brackets denote the expectation
value and $\Delta t_{(rdn)} \equiv t_{(rdn)} - \langle t_{(rdn)} \rangle$.
The $t_{(rdn)}$ are the primary data objects in the harmonic reconstruction
of the microwave sky.

\subsection{The instrument response}
\label{subsec:instrument}

We assume that the detector responses are linear functionals of the sky in the
absence of instrument noise. At any instant, the power seen by a given
detector (due to the sky) is given by convolving the sky with
the appropriate beam and spectral transmission. A single time-ordered data
point is then obtained by convolving this power with the detector's temporal
response.

\subsubsection{Beam patterns}

It is convenient to define the beam profile when the instrument is in the
appropriate horn reference configuration, so that the beams can be assumed not
to vary through the mission. A method for describing arbitrary, polarized
beam patterns in terms of multipoles $b^P_{(lm)}$ was introduced by
Challinor et al.~\shortcite{challinor00}. The beam pattern is first described
in terms of effective Stokes parameters $\{\tilde{I}_d(\be;\nu),
\tilde{Q}_d(\be;\nu), \tilde{U}_d(\be;\nu), \tilde{V}_d(\be;\nu)\}$,
which are defined on the same $\{\bsigma_\theta,-\bsigma_\phi\}$ basis as the
sky. These effective Stokes parameters can be expressed in terms of the
far-field radiation pattern of the detector~\cite{challinor00}.
The $\tilde{Q}_d(\be;\nu)$ and $\tilde{U}_d(\be;\nu)$ can be combined into a
linear polarization tensor for the beam, $\clb_d^{ab}(\be;\nu)$, as in
equation~(\ref{eq:3}). The resultant power per
unit frequency interval, $\ud W_d/\ud\nu$, can then be written in
basis-independent form as the integral
\begin{eqnarray}
\frac{\ud}{\ud\nu}W_d &=& A_{{\rmn{eff}},d}\int[
I(\be;\nu)\tilde{I}_d(\be;\nu)-V(\be;\nu)\tilde{V}_d(\be;\nu) \nonumber \\
&& \mbox{} + 2 \clp_{ab}(\be;\nu) \clb^{ab}_d(\be;\nu)]\, \ud \Omega,
\label{eq:6}
\end{eqnarray}
where $A_{{\rmn{eff}},d}$ is the effective area (assumed independent of
frequency).
The directivity $\tilde{I}_{d}(\be;\nu)$ is a scalar field on the sphere, so
can be expanded in terms of multipoles $b^I_{d(lm)}(\nu)$, as in
equation~(\ref{eq:3}). We impose the constraint
$b^I_{d(00)}(\nu)=1/\sqrt{4\pi}$ so that $\tilde{I}_{d}(\be;\nu)$ integrates
to unity over the sphere. In a similar manner, $\tilde{V}_d(\be;\nu)$
can be represented by multipoles $b^V_{d(lm)}(\nu)$. Finally,
$\clb^{ab}_d(\be;\nu)$ can be expanded in the transverse, trace-free
tensor harmonics with multipoles $b^G_{d(lm)}(\nu)$ and $b^C_{d(lm)}(\nu)$.

To obtain the power per unit frequency interval incident on the spectral filter
chain for the $d$th detector, when the nominal main beam is at angle $\psi$ on
the $r$th ring, we must rotate the beam pattern by
$D(\phi_r,\theta_r,\psi) D(0,\alpha_{rd},\kappa_{rd})$ before performing
the convolution in equation~(\ref{eq:6}). This rotation is easily performed
in multipole space to give the result~\cite{challinor00}
\begin{eqnarray}
\frac{\ud}{\ud\nu}W_{rd}(\psi) &=& A_{{\rmn{eff}},d} \sum_{Plmm'm''}
[a^{P\ast}_{(lm)}(\nu) \tilde{b}^P_{d(lm')}(\nu) \nonumber \\
&& \times D^l_{mm''}(\phi_r,\theta_r,\psi)
D^l_{m'' m'}(0,\alpha_{rd},\kappa_{rd})],
\label{eq:7}
\end{eqnarray}
where, for later convenience, we have defined
\begin{eqnarray}
\tilde{b}^P_{d(lm)}(\nu) &\equiv& \delta_I^P b^I_{d(lm)}(\nu) - \delta^P_V
b^V_{d(lm)}(\nu) \nonumber \\
&&\mbox{} + 2 \delta_G^P b^G_{d(lm)}(\nu) + 2 \delta_C^P b^C_{d(lm)}(\nu),
\label{eq:8}
\end{eqnarray}
with $b^G_{d(lm)}(\nu) = b^C_{d(lm)}(\nu) = 0$ for $l  < 2$.
The $D^l_{mm'}(\alpha,\beta,\gamma)$ appearing in equation~(\ref{eq:7})
are Wigner's $D$-matrices; our conventions follow Brink \&
Satchler~\shortcite{brink93}. The dependence of
$D^l_{mm'}(\alpha,\beta,\gamma)$ on $\alpha$ and $\gamma$ goes as
\begin{equation}
D^l_{mm'}(\alpha,\beta,\gamma) = {\rmn{e}}^{-i m \alpha}
d^l_{mm'}(\beta) {\rmn{e}}^{-i m' \gamma},
\label{eq:9}
\end{equation}
where the $d^l_{mm'}(\beta)$ are the reduced rotation matrices.

\subsubsection{Spectral filtering}
\label{subsubsec:spectra}

The power available at the detection element (e.g. bolometers for high
frequency instruments) is given by integrating $\ud W_{rd}(\psi)/\ud\nu$
against the spectral transmission $v_d(\nu)$ of the filter chain. We shall
assume that the filters on all detectors in a given frequency band are
identical, and have transmission properties that are accurately known
from calibrations on the ground. Map-making from multi-frequency
data is usually performed on a band-by-band basis (although it
is essential to use all the frequency bands to calibrate the
pointing, e.g. van Leeuwen et al.\ 2001). Given that Fourier data
from only detectors in the same frequency band are combined in such an
analysis, we need only consider a single band, and henceforth assume that
$v_d(\nu)$ has no dependence on the detector, i.e. $v_d(\nu) = v(\nu)$.

On integrating equation~(\ref{eq:7}) against $v(\nu)$, the dependence
on the sky enters through the integral $\int a^{P\ast}_{(lm)}(\nu)
\tilde{b}^P_{d(lm')}(\nu) v(\nu) \, \ud\nu$. For the subsequent analysis,
it is important that this integral can  be factored into a part describing the
sky carrying only $l$ and $m$ indices, and a part describing the beam which
has only $l$ and $m'$ indices. Here we assume that the approximate
factorisation
\begin{eqnarray}
\int a^{P\ast}_{(lm)}(\nu)\tilde{b}^P_{d(lm')}(\nu) v(\nu) \, \ud\nu
&\approx& \tilde{b}^P_{d(lm')}(\nu_0) \nonumber \\
&& \times \int a^{P\ast}_{(lm)}(\nu)v(\nu)\,\ud\nu,
\label{eq:10}
\end{eqnarray}
where $\nu_0$ is the central frequency in the band, is accurate to better than
the noise level, so that we can describe the sky by the frequency integrated
multipoles
\begin{equation}
\bar{a}^{P}_{(lm)} \equiv \int a^{P}_{(lm)}(\nu) v(\nu)
\, \ud\nu.
\label{eq:11}
\end{equation}
The largest source of error resulting from this factorisation is expected
to be due to the variation of the width of the main beam across the band.
In Appendix~\ref{app:beam} we compute the systematic error on the
recovered power spectrum of the CMB temperature anisotropies resulting
from this variation, assuming an axisymmetric, diffraction-limited
Gaussian beam, uniform sky coverage, and no foreground contaminants.
For the \emph{Planck} High Frequency Instrument (HFI), the bias may be
non-negligible in the 100 GHz
channel between $l=1000$ and $l=1500$, in which case refinements to
equation~(\ref{eq:10}) will be required. Here, we adopt the factorisation in
equation~(\ref{eq:10}), in which case the power available at the detection
element, $W_{rd}(\psi)$, can be written as
\begin{eqnarray}
W_{rd}(\psi) &=& A_{{\rmn{eff}},d} \sum_{Plmm'}
[\bar{a}^{P\ast}_{(lm)} D^l_{mm'}(\phi_r,\theta_r,\psi) \nonumber \\
&& \phantom{A_{{\rmn{eff}},d} \sum_{Plmm'}} \times \clb^P_{rd(lm')}].
\label{eq:12}
\end{eqnarray}
For later convenience, we have introduced the notation
\begin{equation}
\clb^P_{rd(lm')} \equiv \sum_{m''} D^l_{m'm''}(0,\alpha_{rd},\kappa_{rd})
\tilde{b}^P_{d(lm'')}(\nu_0)
\label{eq:13}
\end{equation}
for the multipoles of the $d$th beam after rotation by
$D(0,\alpha_{rd},\kappa_{rd})$. Note that the $r$-dependence of
$\clb^P_{rd(lm')}$ derives only from the (slow) variation of opening
angle $\alpha_{rd}$ and focal plane rotation $\kappa_{rd}$ through
the mission.

\subsubsection{Temporal response}

The available power, $W_{rd}[\psi(t')]$, is convolved with the temporal
response function of the detector, $h_d(t,t')$, to generate a single
time-ordered datum at time $t$. We decompose $h_d(t,t')$ into a stationary
part, $h_d(t-t')$, and a stochastic part, $\Delta h_d(t,t')$, which we assume
to have stationary statistical properties.
The stochastic part, which derives from e.g.\
random fluctuations in amplifier gains, will be considered as part of the
instrument noise in Section~\ref{subsec:noise}. Here, we concentrate
on the stationary impulse response function $h_d(t)$. Typically, this will
contain several instrument artifacts, such as the effect of finite sampling
and non-zero detector time constants, which are convolved together to
form the total impulse response. For example, if the detector time constant
is $\tau_d$, and samples are taken by integrating over time intervals
$\Delta_d$ and are assigned to the midpoint of the interval,
then $h_d(t) = h_{\tau_d,d}(t) \star
h_{\Delta_d,d}(t)$, where the detector impulse response
\begin{equation}
h_{\tau_d,d}(t) = G_d \tau_d{}^{-1} {\rmn{e}}^{-t/\tau_d} \Theta(t/\tau_d),
\label{eq:14}
\end{equation}
and the sampling impulse response
\begin{equation}
h_{\Delta_d,d}(t) = \Delta_d{}^{-1} \Theta(1+2t/\Delta_d)
\Theta(1-2t/\Delta_d).
\label{eq:15}
\end{equation}
Here, $\Theta(x)$ is the Heaviside unit step function, and $G_d$ is the
average gain over the mission, which includes both amplifier gain and the
quantum efficiency of the detectors.

One of the advantages of describing the data on rings in the Fourier domain
is that the convolution with the temporal response of the instrument reduces
to a simple product~\cite{del98b}. Denoting the contribution of the signal
to the phase-ordered data by $s_{rd}(\psi)$, we have
\begin{equation}
s_{rd}(\psi) = \int_{-\infty}^\infty W_{rd}(\psi + \omega_r t')
h_d(-t')\, \ud t',
\label{eq:16}
\end{equation} 
where $\omega_r = \ud \psi/\ud t$ is the (average) spin rate on the
$r$th ring. We have assumed that the support of $h_d(t)$ is sufficiently
compact that the only contribution to $s_{rd}(\psi)$ is from the available
power $W_{rd}$ on the same ring. We have further assumed that any variation of
the spin rate about the average is negligible for the purpose of performing the
convolution, so that the convolution of the available power with the impulse
response is periodic in $\psi$. Inserting equation~(\ref{eq:12}) into
equation~(\ref{eq:16}), and expanding $D^l_{mm'}(\phi_r,\theta_r,\psi)$ as in
equation~(\ref{eq:9}), we find
\begin{eqnarray}
s_{rd}(\psi) &=& A_{{\rmn{eff}},d} \sum_{Plmm'}
[\bar{a}^{P\ast}_{(lm)} D^l_{mm'}(\phi_r,\theta_r,\psi) \nonumber \\
&& \phantom{A_{{\rmn{eff}},d} \sum_{Plmm'}} \times
H_d^\ast(m' \omega_r) \clb^P_{rd(lm')}],
\label{eq:17}
\end{eqnarray}
where $H_d(\omega_r)$ is the Fourier transform of the impulse response:
\begin{equation}
H_d(\omega_r) = \int_{-\infty}^{\infty} h_d(t) {\rmn{e}}^{-i \omega_r t}
\, \ud t.
\label{eq:18}
\end{equation}
For the impulse responses in equations~(\ref{eq:14}) and~(\ref{eq:15}),
we have
\begin{equation}
H_d(\omega_r) = \frac{G_d\,{\rmn{sinc}}(\omega_r \Delta_d/2)}{1+i\omega_r
\tau_d}.
\label{eq:19}
\end{equation}
Finally, we can extract from equation~(\ref{eq:17}) the contribution of the
signal to the $n$th Fourier coefficient of the phase-ordered data.
Complex conjugating and using equation~(\ref{eq:9}), we find
\begin{equation}
s_{(rdn)} = A_{{\rmn{eff}},d} \sum_{Plm}
[\bar{a}^{P}_{(lm)} d^l_{mn}(\theta_r) {\rmn{e}}^{i m \phi_r}
H_d(n \omega_r) \clb^{P\ast}_{rd(ln)}],
\label{eq:20}
\end{equation}
where the sum is over $l \geq |n|$ and $|m| \leq l$. Note that the
effect of the impulse response can also be described in terms of an effective
(ring-dependent) beam with multipoles
\begin{eqnarray}
b^{P,{\rmn{eff}}}_{rd(lm)}(\nu)&=&\sum_{m'm''} [
D^l_{mm'}(-\kappa_{rd},-\alpha_{rd},0) H_d^\ast(m'\omega_r) \nonumber \\
&& \phantom{\sum_{m'm''} [}
\times D^l_{m'm''}(0,\alpha_{rd},\kappa_{rd}) b^P_{d(lm'')}(\nu)].
\label{eq:21}
\end{eqnarray}
The effective beam will be generally not be axisymmetric as a consequence
of the smearing along the scan direction induced by the impulse response
of the instrument. The concept of an effective beam becomes particularly
significant if the beam has to be calibrated from the observations of
e.g.\ bright point source transits, as described by van Leeuwen et
al.~\shortcite{van01}. In such
cases, it is the effective beam that can be reconstructed most directly.
A potential difficulty that would need to be addressed in such an approach
is the (slow) variation of effective beam with ring due to variations
in the angles $\alpha_{rd}$ and $\kappa_{rd}$. In practice, it may be simpler
to deconvolve the temporal response of the instrument prior to construction
of the phase-ordered data. In this case, $H_d(n\omega_r)$ can be omitted from
equation~(\ref{eq:20}).

\subsection{Gaussian random noise and reconstruction errors}
\label{subsec:noise}

The Fourier modes, $t_{(rdn)}$, that we construct from the phase-ordered data
differ from the $s_{(rdn)}$ of equation~(\ref{eq:20}) because of a number
of sources of error. For example, stochasticity in the instrument temporal
response, such as that due to thermal noise in the detectors and
amplifiers, has not been included in equation~(\ref{eq:20}). Neither have we
included the effects of photon (shot) noise, or any detector offsets.
Furthermore, we have replaced the spin axis pointing by its average
over the $N_{\rmn{s}}$ spin periods, ignoring the (albeit small) effects of the
nutation of the instrument. These random errors tend to be suppressed in the
phase-ordered data due to the averaging over $N_{\rmn{s}}$ periods, except
for fluctuations at temporal frequencies synchronous with the spin frequency.
Fluctuations below the spin frequency remain
in the phase-ordered data predominantly as an offset on the ring. In the
usual map-making paradigm~\cite{bond99,borrill99}, low frequency ($1/f$)
noise must be carefully accounted for; failure to do so leads to highly
correlated errors in the reconstructed map which typically appear as
stripes (e.g.\ Delabrouille 1998; Revenu et al.\ 2000). It is difficult to
accommodate low frequency noise power in map-making from the time-ordered data
since its inclusion spoils the sparse nature of the time-time noise
correlation matrix. Approximate methods to circumvent this problem for
experiments that scan on rings already exist~\cite{del98a,revenu00}; these
`destriping' methods will be discussed further in Section~\ref{subsec:stripes}.

There are also a number of potential systematic errors which
contaminate the $t_{(rdn)}$, and, if left unaccounted for, will lead to bias
in the reconstructed multipoles of the sky. Keeping the numerous systematics
under control is essential to maintaining the integrity of the final data
products. A significant merit of the harmonic data model is that
the model of the instrument, equation~(\ref{eq:20}), is sufficiently complete
to include a number of systematic effects which are not so naturally included
in the standard map-making techniques. For example, the harmonic model can
seamlessly accommodate any prior knowledge we have of asymmetries in the beam
profiles, cross-polar contamination for the polarized detectors, sidelobe
features leading to `straylight' entering the instrument, and detector
time constants and data sampling rate. For some systematics there will be
incomplete (or even no) prior knowledge of essential parameters, in which
case these parameters must be estimated from the data themselves during
the reconstruction of the sky. The harmonic model also appears
well-suited to handling such systematics iteratively; see
Section~\ref{subsec:sync}.

To include the effect of Gaussian random noise with a known (or estimated)
power spectrum in the reconstruction of the sky from the Fourier data
on rings, $t_{(rdn)}$, we require the covariance matrix
$N_{(rdn)(r'd'n')}$. This matrix
was calculated by Delabrouille at al.~\shortcite{del98b} for a single
detector and ring ($d=d'$ and $r=r'$). Here we extend their result
to include the effects of a finite sampling rate (see also
Janssen et al.\ 1996), and to allow for $r\neq r'$. Assuming instrument noise
dominates the random errors $\Delta t_{(rdn)}$, there will be negligible
correlation between the errors on different detectors, so we need only
consider $d=d'$.

We assume that the noise contribution, $n_d(t)$, to the time stream of the
$d$th detector is a real, stationary random process with zero mean, and power
spectrum $\cln_d(\omega)$. The power spectrum is the Fourier transform
of the noise auto-correlation function $C_d(\tau) = \langle n_d(t+\tau)
n_d(t) \rangle$:
\begin{equation}
\cln_d(\omega) = \int_{-\infty}^\infty C_d(\tau) {\rmn{e}}^{-i\omega\tau}
\, \ud \tau.
\label{eq:22}
\end{equation}
The propagation of the noise $n_d(t)$ to the errors $\Delta t_{(rdn)}$
depends on the exact procedure for estimating the Fourier modes $t_{(rdn)}$
from the time-ordered data~\cite{van01}. However, to a first approximation,
the errors $\Delta t_{(rdn)}$ are obtained by convolving $n_d(t)$ with
the impulse response of the sampler, $h_{\Delta_d,d}(t)$, mapping the
portion of this function covering the observation interval for the $r$th ring
onto the angle $\psi$ around the ring, and finally extracting the Fourier
coefficients of the resultant signal:
\begin{equation}
\Delta t_{(rdn)}=\frac{\omega_r}{2\pi}\int_{t_{rd}^{-}}^{t_{rd}^{+}}
h_{\Delta_d,d} \star n_d(t) 
\exp[-in \omega_r (t-t_{rd}^{-})] \, \ud t.
\label{eq:23}
\end{equation}
Here, $t_{rd}^{-}$ is the time when the $d$th detector is first
pointing to $\psi=0$ on the $r$th ring, $t_{rd}^{+}$ is the end of the
observation interval for that ring:
$t_{rd}^{+}=t_{rd}^{-}+2\pi N_{\rmn{s}} /\omega_r$,
and we have ignored any variation of the spin rate during the $N_{\rmn{s}}$
revolutions. It is now straightforward to write the covariance of the errors
in terms of the noise power spectrum. For the case of scanning missions
such as \emph{Planck}, the number of revolutions $N_{\rmn{s}} \gg 1$ is
sufficiently large that negligible correlations remain between
different Fourier modes. In this case we find
\begin{eqnarray}
N_{(rdn)(r'd'n')} &=&
\delta_{dd'}\delta_{nn'}\frac{1}{2\pi}\int_{-\infty}^\infty \{ \cln_d(\omega)
{\rmn{sinc}}^2 (\omega \Delta_d/2) \nonumber \\
&&\mbox{} \times
{\rmn{sinc}}^2[(\omega/\omega_r-n)\pi N_{{\rmn{s}}}] \nonumber \\
&&\mbox{} \times \exp[i2\pi N_{\rmn{s}} (r-r')\omega/\omega_r] \}
\, \ud\omega,
\label{eq:24}
\end{eqnarray}
where we have further ignored any variation of the average spin rate,
$\omega_r$, between rings. Provided that $\cln_d(\omega){\rmn{sinc}}^2
(\omega \Delta_d/2)$ varies slowly compared to the rest of the integrand
in equation~(\ref{eq:24}) around $\omega=n\omega_r$, we can replace it by its
value there. The remaining integral vanishes unless $r=r'$,
so that for slowly varying power spectra the covariance matrix is fully
diagonal:
\begin{eqnarray}
N_{(rdn)(r'd'n')} &=& \delta_{rr'}\delta_{dd'}\delta_{nn'} \nonumber \\
&&\mbox{}\times
\frac{\omega_r}{2\pi N_{\rmn{s}}} \cln_d(n \omega_r)
{\rmn{sinc}}^2(n\omega_r \Delta_d/2).
\label{eq:25}
\end{eqnarray}

In the presence of a significant low frequency noise component (such as
$1/f$ noise in the electronics) joining onto white noise at a knee
frequency $\omega_{\rmn{knee}}$ (e.g. Delabrouille 1998), the assumption
of a slowly varying noise power spectrum may not hold for small $|n|$.
Typically the spin rate will be chosen so that $\omega_{\rmn{knee}} \ll
\omega_r$, in which case equation~(\ref{eq:25}) will be valid for $n \neq 0$.
However, for $n=0$ we can expect correlations between rings unless
$\omega_r/N_{{\rmn{s}}} \ll \omega_{\rmn{knee}} \ll \omega_r$.
(For the \emph{Planck} HFI, the nominal $\omega_{\rmn{knee}} < 0.06 \,
{\rmn{rad}}\,{\rmn{s}}^{-1}$, and $\omega_r \approx 0.10\, {\rmn{rad}}\,
{\rmn{s}}^{-1}$, so these conditions are likely to be satisfied.)
Even in the presence of such correlations, the noise covariance matrix is
still block diagonal in the harmonic representation, in contrast to the
time-time covariance matrix which has significant off-diagonal terms due to
the extended support of the noise auto-correlation function, $C_d(\tau)$.

\subsection{Maximum-likelihood solution}
\label{subsec:ml}

Our model for the ring Fourier data $t_{(rdn)}$ is now $t_{(rdn)} =
s_{(rdn)} + \Delta t_{(rdn)}$, where the signal $s_{(rdn)}$ is modelled
by equation~(\ref{eq:20}). It is convenient to write this relation in the form
\begin{equation}
t_{(rdn)} = \sum_{Plm} A_{(rdn)(Plm)} \bar{a}^P_{(lm)} + \Delta t_{(rdn)},
\label{eq:26}
\end{equation}
where the coupling matrix is
\begin{equation}
A_{(rdn)(Plm)} \equiv  A_{{\rmn{eff}},d} d^l_{mn}(\theta_r)
{\rmn{e}}^{i m \phi_r} H_d(n \omega_r) \clb^{P\ast}_{rd(ln)}.
\label{eq:27}
\end{equation}
We define $A_{(rdn)(Plm)}$ to be zero for $l < |n|$, since the
$t_{(rdn)}$ only depend on the $\bar{a}^P_{(lm)}$ with $l \geq |n|$.
This structure can be exploited to derive an unbiased (although not minimal
variance) estimator for the $\bar{a}^P_{(lm)}$, which can be computed
in $O(l_{\rmn{max}}^4)$ operations. Here, $l_{\rmn{max}}$ is the maximum
$l$ that we retain during the inversion. (The experimental beam limits
$l_{\rmn{max}}$ even if the underlying sky is not band-limited.) The estimate
is derived by working down from the maximum Fourier mode, $n_{\rmn{max}}$,
and performing a regularised inversion of a subset of the data to solve
for those $\bar{a}^P_{(lm)}$ on which the data subset depends but which have
not been determined at a previous step. Full details of this estimator
will be given in a future paper (Mortlock et al.\ in preparation).

In this paper we shall just give the formal maximum-likelihood solution
to equation~(\ref{eq:26}). We assume that we are attempting to solve only
for the $\bar{a}^P_{(lm)}$ which form the components of a vector $\bmath{a}$.
In practice, it may also be desirable to attempt to solve for
several instrument parameters which are unknown from earlier calibrations,
but which represent significant systematic effects. Such parameters are
easily included in the maximum-likelihood formalism, but unless they influence
the data linearly their inclusion would prevent us from being able to locate
the maximum-likelihood multipoles analytically. Given the size of the
problem, having to locate the maximum of the likelihood by a numerical search
in model space is clearly undesirable. The treatment of systematics
with unknown parameters is probably best performed iteratively
(e.g. Delabrouille, Gispert \& Puget 1998b;
see also Section~\ref{subsec:sync}).

Assuming Gaussian noise, the maximum-likelihood estimate, $\hat{\bmath{a}}$,
for the true sky, $\bmath{a}$, is
\begin{equation}
\hat{\bmath{a}} = ({\mathbfss{A}}^\dagger{\mathbfss{N}}^{-1}
{\mathbfss{A}})^{-1} {\mathbfss{A}}^\dagger {\mathbfss{N}}^{-1} \bmath{t},
\label{eq:28}
\end{equation}
where $\bmath{t}$ is the vector of Fourier modes $t_{(rdn)}$,
$\mathbfss{A}$ is the matrix of coupling coefficients
$A_{(rdn)(r'd'n')}$, and $\mathbfss{N}$ is the noise covariance matrix
with components $N_{(rdn)(r'd'n')}$. The ${}^\dagger$ operation in
equation~(\ref{eq:28})  denotes Hermitian conjugation. The maximum-likelihood
solution is the optimal, unbiased estimate of the sky. Its covariance matrix
is
\begin{equation}
{\mathbfss{C}} \equiv \langle (\hat{\bmath{a}} - \bmath{a})
(\hat{\bmath{a}} - \bmath{a})^\dagger \rangle =
({\mathbfss{A}}^\dagger{\mathbfss{N}}^{-1}{\mathbfss{A}})^{-1},
\label{eq:29}
\end{equation}
where the average is over noise realisations. A brute force evaluation
of $\hat{\bmath{a}}$ requires $O(l_{\rmn{max}}^6)$ operations and
$O(l_{\rmn{max}}^4)$ storage which is
impossible at the moment for high resolution experiments.
Conjugate gradient techniques
(e.g. Press et al.\ 1992) would reduce the operations count to
$O(N_{\rmn{i}}l_{\rmn{max}}^4)$, where $N_{\rmn{i}}$ is the number of
iterations required, by removing the matrix multiplies
and inversion. Note that the block-diagonal structure of
$\mathbfss{N}$ (equation~\ref{eq:24}) reduces the cost of computing
$\mathbfss{N}^{-1}$ to $O(l_{\rmn{max}}^4)$ operations, or fewer
if correlations between rings are confined to the $n=0$ modes.
Keeping $N_{\rmn{i}}$ small requires a careful choice
of preconditioner, i.e.\ a Hermitian, positive-definite matrix which is both
a good approximation to the inverse covariance matrix ${\mathbfss{C}}^{-1}$
and is easy to invert. We demonstrate
in Section~\ref{sec:scan} that for simple scan strategies such matrices
can easily be found (see also Oh et al.\ 1999). For the case where
all rings are at similar latitude, an approximation to
${\mathbfss{C}}^{-1}$ can be computed
in $O(l_{\rmn{max}}^4)$ operations and requires only $O(l_{\rmn{max}}^3)$
storage. This preconditioner is block-diagonal and can be inverted in
$O(l_{\rmn{max}}^4)$ operations. Note, however, that the conjugate gradient
evaluation of equation~(\ref{eq:29}) still requires $O(l_{\rmn{max}}^4)$
storage for the non-sparse matrix $\mathbfss{A}$. The application of the
conjugate gradient method, as well other iterative techniques, to
equation~(\ref{eq:29}) will be explored numerically in
Mortlock et al.~(in preparation).

In writing down the maximum-likelihood solution we have assumed that
${\mathbfss{A}}^\dagger{\mathbfss{N}}^{-1}\mathbfss{A}$ is invertible, so that
the solution is unique. For some scan strategies and instrument geometries,
${\mathbfss{A}}^\dagger{\mathbfss{N}}^{-1}{\mathbfss{A}}$ may become
numerically
singular due to incomplete coverage of the sky. With incomplete coverage,
and $l_{\rmn{max}}$ sufficiently large, there may exist sets
of multipoles which produce a temperature or polarization field which is
localised to machine precision in the regions of the sky that are not
covered~\cite{mort01}. In such cases, solving for the multipoles must
proceed via singular value techniques, and we obtain no constraint on the
part of $\bmath{a}$ which lies in the null space of the coupling matrix
$\mathbfss{A}$. If map-making were performed in pixel space, similar problems
would arise when attempting to estimate the multipoles from the pixelised
map. As an alternative to singular-value techniques, we could further
regularise the inversion by Wiener filtering. With a Gaussian prior
for the signal $\bmath{a}$, with covariance
${\mathbfss{S}}\equiv \langle \bmath{a}\bmath{a}^\dagger \rangle$, the
maximum of the posterior probability gives the Wiener-filtered reconstruction
\begin{equation}
\hat{{\bmath{a}}} = ({\mathbfss{S}}^{-1} + {\mathbfss{A}}^\dagger
{\mathbfss{N}}^{-1} {\mathbfss{A}})^{-1} {\mathbfss{A}}^\dagger
{\mathbfss{N}}^{-1} {\bmath{t}}.
\label{eq:29a}
\end{equation}
In this manner, the inversion of the singular matrix
${\mathbfss{A}}^\dagger{\mathbfss{N}}^{-1} {\mathbfss{A}}$ is
regularised by the inverse signal covariance matrix ${\mathbfss{S}}^{-1}$.
In practice, the microwave sky is only poorly approximated as a Gaussian
random process, particularly in those frequency channels where the CMB is not
dominant. In addition, we will only have limited knowledge of the signal
covariance matrix ${\mathbfss{S}}$ either from other observations, or a
preliminary, approximate analysis of the data. However, Wiener filtering
has proved to be a useful technique in component separation where similar
objections to the use of Gaussian priors hold (see
e.g. Hobson et al.\ 1998 and references therein), and the same should be true
for harmonic `map'-making.
Note that Wiener filtering produces a biased estimate of the
sky; other linear filters have been proposed to cirumvent this problem, but
generally produce noisier, though unbiased, maps (e.g.\ Tegmark \&
Efstathiou 1996).
Equation~(\ref{eq:29a}) can also be solved with conjugate gradient
techniques; in this case the preconditioner should include some
simplified form of the signal covariance in addition to an approximation
to ${\mathbfss{A}}^\dagger{\mathbfss{N}}^{-1} {\mathbfss{A}}$ .

In our discussion so far we have assumed that full beam deconvolution is
performed during the
`map'-making stage. In practice, deconvolving the beam completely may be
undesirable for several reasons. Firstly, the matrix 
${\mathbfss{A}}^\dagger{\mathbfss{N}}^{-1} {\mathbfss{A}}$ is
likely to be more ill-conditioned since the errors on the reconstructed
multipoles
must grow large as $l$ approaches the (inverse) beam scale. (In a conjugate
gradients solution, the preconditioner can partly take this effect into
account.) Secondly, as discussed in Section~\ref{subsubsec:spectra} and
Appendix~\ref{app:beam}, variation of the width of the main beam across
the spectral band of a channel is a potential source of systematic error.
For an asymmetric beam, one could make the optimistic assumption
beam multipoles in a given frequency channel could be written as the
product of a frequency-dependent, symmetric part $\tilde{b}^P_{(l0)}(\nu)$
which is assumed to be the same for all detectors in the band, and a
frequency-independent, asymmetric part which may depend on the detector:
\begin{equation}
\tilde{b}^P_{d(lm)}(\nu) \approx \tilde{b}^P_{(l0)}(\nu)
[\tilde{b}^P_{d(lm)}(\nu_0)/ \tilde{b}^P_{(l0)}(\nu_0)],
\end{equation}
where $\nu_0$ is the central frequency in the band. We could then solve for
the multipoles of the sky convolved with an effective symmetric beam with
multipoles
$\tilde{b}^P_{(lm)}(\nu)$ and integrated across the frequency band with
filter $v(\nu)$. By factoring the beam in this way, we can simultaneously
improve the condition number of ${\mathbfss{A}}^\dagger{\mathbfss{N}}^{-1}
{\mathbfss{A}}$ and remove most of the bias due to variation of the main beam
width with frequency.

\section{Simple scan strategies}
\label{sec:scan}

In this section we investigate the structure of the inverse covariance
matrix, $\mathbfss{C}^{-1}$, of the recovered multipoles for two simple
scanning strategies. The first is a random pointing of the spin axis for
each ring, which ensures uniform coverage of the entire sky. The second
is constant latitude scanning, where all rings have the same polar angle,
$\theta_r = \theta$. Constant latitude scanning is a useful
approximation to the scan strategy of the \emph{Planck} satellite, where
the spin axis stays close to the plane of the ecliptic. By adopting
coarse restrictions on the behaviour of the instrument, we are able to
demonstrate that, under these restrictions, the harmonic estimate of the
sky, equation~(\ref{eq:28}), is statistically equivalent to that obtained
via conventional map-making and spherical analysis. By relaxing some
restrictions on the behaviour of the instrument, we obtain
some useful extensions of the known results.

Throughout this section we impose a number of limits on the
instrument response: (i) The spin axis position is always
aligned with the $z$-axis of the instrument reference system (see
Section~\ref{subsec:ring}), so that the ring opening angles, $\alpha_{rd}$,
are independent of $r$, and the focal plane rotations $\kappa_{rd}=0$;
(ii) There is no variation in average spin rate between
rings, so we can drop the subscript from $\omega_r$;
(iii) The band width of the spectral filters is
sufficiently narrow that
\begin{eqnarray}
\int a^{P\ast}_{(lm)}(\nu)\tilde{b}^P_{(lm')}(\nu) v(\nu)\,\ud \nu
&=& a^{P\ast}_{(lm)}(\nu_0)\tilde{b}^P_{(lm')}(\nu_0) \nonumber \\
&&\mbox{} \times \int v(\nu)\, \ud\nu,
\label{eq:31}
\end{eqnarray}
where $\nu_0$ is the central frequency of the band; (iv) Bolometer time
constants, $\tau_d$, and the sampling period, $\Delta_d$, are such that
$\tau_d \omega_r l_{\rmn{max}}\ll 1$ and
$\Delta_d \omega_r l_{\rmn{max}}\ll 1$, in which case the effect of the
instrument response can be approximated by simply a gain at zero lag;
(v) The noise power spectrum is white so that
\begin{equation}
N_{(rdn)(r'd'n')} = \delta_{rr'} \delta_{dd'} \delta_{nn'}
\sigma_d^2,
\label{eq:32}
\end{equation}
where $\sigma^2_d$ is a constant for each detector, which can be
related to the detector sensitivity, $s_d$.\footnote{The detector
sensitivity is defined such that for
an integration time $\Delta t$ with a single pointing, the excess signal to
noise is unity when the instrument is illuminated with an unpolarized
black-body sky with a temperature excess
$\bar{T} s/ \sqrt{\Delta t}$ over the average CMB
temperature, $\bar{T}$. If the units of the gain $G_d$ are e.g.\ $\mbox{V}
\mbox{W}^{-1}$, so that $s_{rd}(\psi)$ is a readout Voltage, $\sigma_d$
will also have the dimensions of a Voltage, and $s_d$ will have units of
e.g.\ $(\mu\mbox{K}/\mbox{K})\mbox{s}^{1/2}$.}
With the restrictions above,
\begin{eqnarray}
s_d &=& \left[A_{{\rmn{eff}},d} G_d \bar{T}
\left. \frac{\partial B(\nu_0,T)}{\partial T} \right|_{T=\bar{T}}
\int v(\nu) \, \ud\nu\right]^{-1} \nonumber \\
&&\mbox{} \times \sqrt{\frac{2\pi N_{\rmn{s}}}{\omega_r}} \sigma_d, 
\label{eq:33}
\end{eqnarray}
where $B(\nu,T)$ is the Planck function and the average CMB temperature is
$\bar{T}$. Note that a consequence of restriction (i) is that $\clb^P_{rd(ln)}$
is independent of ring, $\clb^P_{rd(ln)}=\clb^P_{d(ln)}$. If we define
dimensionless multipoles by
\begin{equation}
\tilde{a}^P_{(lm)} \equiv \left[\bar{T} \left.
\frac{\partial B(\nu_0,T)}{\partial T} \right|_{T=\bar{T}}
\right]^{-1} a^P_{(lm)}(\nu_0), 
\label{eq:34}
\end{equation}
the inverse of the dimensionless error covariance matrix,
$\tilde{\mathbfss{C}}^{-1}$, has components
\begin{eqnarray}
\tilde{C}^{-1}_{(Plm)(P'l'm')} &=& \frac{2\pi N_{\rmn{s}}}{\omega}\sum_{rdn}
s_d^{-2}\{ D^l_{mn}(\phi_r,\theta_r,0)\clb^{P}_{d(ln)} \nonumber \\
&&\mbox{} \times [D^{l'}_{m'n}(\phi_r,\theta_r,0)
\clb^{P'}_{d(l'n)}]^\ast \},
\label{eq:35}
\end{eqnarray}
where the sum is over detectors, rings, and $|n| \leq {\rmn{min}}(l,l')$.
We now proceed to analyse $\tilde{\mathbfss{C}}^{-1}$ for randomly-positioned
rings and constant latitude scanning.

\subsection{Randomly-positioned rings}
\label{subsec:random}

We consider the limit of equation~(\ref{eq:35}) as the number of rings
$N_{\rmn{r}} \rightarrow \infty$ while keeping the survey length finite,
in which case we can replace the sum over rings by an appropriate integral.
In practice, for finite $N_{\rmn{r}}$ the continuum limit should hold for
scales with $l \la \sqrt{N_{\rmn{r}}/(4\pi)}$. Placing the rings at random
is equivalent to enforcing uniform coverage of the full sky, which will
allow us to make contact with existing analytic results obtained from
the map-based formalism. The sum over randomly-positioned rings in
equation~(\ref{eq:35}) can be replaced by $[N_{\rmn{r}}/(4\pi)]\int
\, \ud\Omega_r$, where $\ud\Omega_r \equiv \ud\phi_r \,\ud \cos\theta_r$.
Making use of the orthogonality of the
$D^l_{mm'}(\alpha,\beta,\gamma)$ over the ${\rmn{SO}}(3)$ group
manifold~\cite{brink93}, we find
\begin{equation}
\int D^l_{mn}(\phi_r,\theta_r,0) [D^{l'}_{m'n}(\phi_r,\theta_r,0)]^\ast
\,\ud\Omega_r = \frac{4\pi}{2l+1} \delta_{ll'} \delta_{mm'},
\label{eq:36}
\end{equation}
so that $\tilde{\mathbfss{C}}^{-1}$ simplifies to
\begin{equation}
\tilde{C}^{-1}_{(Plm)(P'l'm')} =\delta_{ll'}\delta_{mm'}
\frac{T_{\rmn{m}}}{2l+1} \sum_{dn} s_d^{-2} \clb^{P}_{d(ln)}
\clb^{P'\ast}_{d(ln)},
\label{eq:37}
\end{equation}
where the total mission time is $T_{\rmn{m}} = 2\pi N_{\rmn{s}} N_{\rmn{r}}/
\omega$. Equation~(\ref{eq:37}) can be further simplified by using
equation~(\ref{eq:13}) to substitute for $\clb^{P}_{d(ln)}$ and then
performing the sum over $n$ noting that $[D^l_{mn}(\alpha,\beta,\gamma)]^\ast
= D^l_{nm}(-\gamma,-\beta,-\alpha)$. The result is
\begin{eqnarray}
\tilde{C}^{-1}_{(Plm)(P'l'm')} &=& \delta_{ll'}\delta_{mm'}
\frac{T_{\rmn{m}}}{2l+1} \nonumber \\
&&\mbox{} \times \sum_{dm''} s_d^{-2} \tilde{b}^{P}_{d(lm'')}(\nu_0)
\tilde{b}^{P'\ast}_{d(lm'')}(\nu_0).
\label{eq:38}
\end{eqnarray}
Note that there is now no dependence on the opening angles of the detectors,
$\alpha_{rd}=\alpha_d$. This is a consequence of assuming white
noise. If the noise had a characteristic timescale, this would combine with
the spin velocity and opening angle to define a characteristic angular scale
for the projected noise, and $\tilde{\mathbfss{C}}$ would then depend on
$\alpha_d$.

To make further progress, we consider initially the case where
the beams are axisymmetric, and have no cross-polar contamination.
In this limit, the intensity multipoles can be written in terms of a
window function, $W_{l,d}$:
\begin{equation}
b^I_{d(lm)}(\nu_0) = \sqrt{\frac{2l+1}{4\pi}} W_{l,d} \delta_{m0},
\label{eq:39}
\end{equation}
and the linear polarization multipoles in terms of spin-weight 2 window
functions, ${}_2W_{l,d}$, (Challinor et al.\ 2000; see also Ng \& Liu
1999):
\begin{eqnarray}
b^G_{d(lm)}(\nu_0) &=&- \sqrt{\frac{2l+1}{32\pi}}{}_2W_{l,d}
{\rmn{e}}^{-im\rho_d}(\delta_{m2}+\delta_{m\, -2}), \label{eq:40} \\
b^C_{d(lm)}(\nu_0) &=&-i\sqrt{\frac{2l+1}{32\pi}}{}_2W_{l,d}
{\rmn{e}}^{-im\rho_d}(\delta_{m2}-\delta_{m\, -2}). \label{eq:41}
\end{eqnarray}
Here, $\rho_d$ is the angle between the polarization direction on axis
and the normal to the plane containing the main beam direction and the
spin axis. (The unit normal to this plane is $\bsigma_y$ when the detector is
in its horn's reference configuration). By assumption, the circular
polarization multipoles for the beam vanish.
Substituting into equation~(\ref{eq:38}), we find
\begin{eqnarray}
\tilde{C}^{-1}_{(Plm)(P'l'm')} &=& \delta_{ll'}\delta_{mm'}\delta_{PP'}
\nonumber \\
&&\hspace{-10pt}\times \sum_d w_d [\delta^P_I W_{l,d}^2
+ (\delta^P_G + \delta^P_C){}_2W_{l,d}^2],
\label{eq:42}
\end{eqnarray}
where we have introduced the weight per solid angle~\cite{knox95}, which
for uniform coverage of the sky is $w_d \equiv T_{\rmn{m}}/(4\pi s_d^2)$.
Note that the covariance matrix is diagonal, and that it does not depend on
the relative orientations of the polarization directions of the various
detectors. This is not surprising since for randomly-positioned rings, every
point on the sky is traversed by every detector in every possible orientation,
and we have assumed the instrument noise to be uncorrelated between detectors.

It is straightforward to show that the covariance matrix in
equation~(\ref{eq:42}) is equal to that obtained by constructing
optimal (minimum variance), beam-smoothed maps of the Stokes parameters
for each detector, and then optimally extracting the $\tilde{a}^P_{(lm)}$ from
a joint analysis of the maps. If the main beam of the $d$th detector lies in
the $p$th pixel when the ring phase is $\psi$ on the $r$th ring, the signal
contribution to the phase-ordered data, equation~(\ref{eq:17}), can be
written as (see e.g.\ Challinor et al. 2000)
\begin{eqnarray}
s_{rd}(\psi) &=& A_{{\rmn{eff}},d}G_d \bar{T}\left.
\frac{\partial B(\nu_0,T)}{\partial T}\right|_{T=\bar{T}}
\int v(\nu) \, \ud\nu [I_{{\rmn{eff}},d}(\be_p) \nonumber \\
&&\mbox{}- Q_{{\rmn{eff}},d}(\be_p) \cos 2\eta
+ U_{{\rmn{eff}},d}(\be_p) \sin 2\eta].
\label{eq:43}
\end{eqnarray}
under the restrictions described above. In equation~(\ref{eq:43}),
$\be_p$ is in the direction of the $p$th pixel, and has polar angle
$\theta_p$ and azimuth $\phi_p$, and the angle $\eta$ is the angle between
the polarization direction (on the sky) along the beam axis and the
plane containing $\be_p$ and the $z$-axis of the fixed reference system.
The beam-smoothed fields are
\begin{eqnarray}
I_{{\rmn{eff}},d}(\be) &=& \sum_{lm} W_{l,d} \tilde{a}^I_{(lm)} Y_{(lm)}(\be)
\label{eq:44} \\
\frac{1}{\sqrt{2}}(Q_{{\rmn{eff}},d}\pm i U_{{\rmn{eff}},d})(\be)
&=& \sum_{lm}[{}_2W_{l,d} (\tilde{a}^G_{(lm)}\mp i\tilde{a}^C_{(lm)})
\nonumber \\
&&\mbox{} \times {}_{\mp 2}Y_{(lm)}(\be)],
\label{eq:45}
\end{eqnarray}
where the ${}_{\pm 2}Y_{(lm)}$ are examples of the spin-weight harmonics,
defined for integer $s$ by~\cite{goldberg67}
\begin{equation}
D^l_{m\, -s}(\alpha,\beta,\gamma) = (-1)^s \sqrt{\frac{4\pi}{2l+1}}
{}_s Y^\ast_{(lm)}(\beta,\alpha) {\rmn{e}}^{is\gamma}.
\label{eq:46}
\end{equation}
For randomly-positioned rings in the limit $N_{\rmn{r}} \rightarrow \infty$,
the angles $\psi$ are uniformly covered in any given pixel. Given our
assumptions, the optimal, unbiased estimate for $I_{{\rmn{eff}},d}$ in the
$p$th pixel is given by direct averaging of the appropriate $t_{rd}(\psi)$. For
$Q_{{\rmn{eff}},d}$ the $t_{rd}(\psi)$ are averaged with weight $-2
\cos 2\eta$, and for $U_{{\rmn{eff}},d}$ the weight is $2\sin 2\eta$.
The errors on the smoothed maps are uncorrelated between Stokes parameters,
and between pixels. The errors on these maps have weights per solid angle
$w_d$ for $I_{{\rmn{eff}},d}$, and $w_d/2$ for $Q_{{\rmn{eff}},d}$ and
$U_{{\rmn{eff}},d}$.

For uncorrelated noise between Stokes parameters, pixels, and detectors,
the maximum-likelihood estimate of the $\tilde{a}^P_{(lm)}$ from the smoothed
maps reduces to minimising the (absolute) squared residuals between the left
and right-hand sides of equations~(\ref{eq:44}) and~(\ref{eq:45}), with
each pixel for each detector entering with the appropriate weight per
solid angle. In the case of uniform coverage considered in this subsection,
the maximum-likelihood estimates of the $\tilde{a}^P_{(lm)}$ are equivalent to
performing a spherical transform of the smoothed maps and then averaging
these across the detectors, with each detector carrying statistical weight
$W_{l,d}w_d$. However, it will be useful for the next subsection to note
the form that the error covariance matrix takes when we relax the assumption
of uniform coverage, while retaining the assumptions of uncorrelated errors
between Stokes parameters, detectors and pixels. The weights per solid
angle for $I_{{\rmn{eff}},d}$ are then pixel-dependent, $w_{d,p}$, and
the weights for $Q_{{\rmn{eff}},d}$ and $U_{{\rmn{eff}},d}$ are $w_{d,p}/2$.
The non-vanishing entries of the inverse of the (Hermitian) error covariance
matrix are then:
\begin{eqnarray}
\tilde{C}^{-1}_{(Ilm)(Il'm')} &=& \sum_{pd} W_{l,d}W_{l',d}w_{d,p}
Y_{(lm)}^\ast Y_{(l'm')} \Delta \Omega_p, \label{eq:47} \\
\tilde{C}^{-1}_{(Glm)(Gl'm')} &=& \sum_{pd} [ {}_2W_{l,d}\, {}_2W_{l',d}w_{d,p}
{\textstyle{\frac{1}{2}}}({}_2Y_{(lm)}^\ast {}_2Y_{(l'm')} \nonumber \\
&&\phantom{\sum_{pd}} +{}_{-2}Y_{(lm)}^\ast{}_{-2}Y_{(l'm')})
\Delta \Omega_p], \label{eq:48} \\
\tilde{C}^{-1}_{(Glm)(Cl'm')} &=& i\sum_{pd} [{}_2W_{l,d}\, {}_2W_{l',d}w_{d,p}
{\textstyle{\frac{1}{2}}}({}_2Y_{(lm)}^\ast {}_2Y_{(l'm')} \nonumber \\
&&\phantom{\sum_{pd}} - {}_{-2}Y_{(lm)}^\ast{}_{-2}Y_{(l'm')})
\Delta \Omega_p], \label{eq:49}
\end{eqnarray}
and $\tilde{C}^{-1}_{(Clm)(Cl'm')}=\tilde{C}^{-1}_{(Glm)(Gl'm')}$. Here,
$\Delta \Omega_p$ is the solid angle subtended by the $p$th pixel. For the
case of uniform coverage, $w_{d,p}=w_d$, the orthonormality of the
spin-weight harmonics reduces equations~(\ref{eq:47})--(\ref{eq:49})
to the result derived from the harmonic model, equation~(\ref{eq:42}), if
we ignore pixelisation effects. For this simple example, the estimates
of the $a^P_{(lm)}$ from the harmonic model and the map-making route
are both unbiased and have the same error covariance. It follows that they
are statistically equivalent at the level of second-order correlations
(and at all orders for Gaussian noise).

\subsubsection{Effect of beam asymmetry}

We now consider the effect of a known beam asymmetry on the covariance
matrices obtained via the harmonic and map-based routes. For simplicity,
we shall only consider unpolarized detectors with bivariate Gaussian
profiles, with eccentricity $e_d$. Consider
initially a detector in its horn's reference configuration, with
the major axis of the iso-directivity ellipse aligned with $\bsigma_x$.
For beam widths $\sigma_d \ll 1$, we have
\begin{equation}
\tilde{I}_d(\be;\nu_0) \approx \frac{\exp\{-\theta^2
[\cos^2\phi + \sin^2\phi/(1-e_d^2)]/(2\sigma_d^2)\}}
{2\pi\sigma_d^2\sqrt{1-e_d^2}},
\label{eq:50}
\end{equation}
which has non-zero multipoles, in the limit of large $l$,
\begin{equation}
b^I_{d(lm)} = \sqrt{\frac{2l+1}{4\pi}}
I_{m/2}(l^2 \sigma^2 e_d^2/4){\rmn{e}}^{-l^2\sigma_d^2(1-e_d^2/2)/2},
\label{eq:51}
\end{equation}
for $m$ even, where $I_{m}(x)$ are modified Bessel functions. Note that
beam asymmetry is only significant for $l\gg 1/(\sigma_d e_d)$. The multipoles
for a detector whose iso-directivity ellipse has major axis at
angle $\gamma_d$ to $\bsigma_x$ pick up an additional phase
$\exp(-im\gamma_d)$.

The expression for $\tilde{C}^{-1}_{(Ilm)(Il'm')}$ (equation~\ref{eq:38})
involves the sum $\sum_{|n| \leq l} |b^I_{d(ln)}(\nu_0)|^2$. For large $l$,
this is easily evaluated by substituting the integral representation of the
modified Bessel functions (or, equivalently, azimuthally averaging the
absolute square of the flat-space Fourier transform of the directivity). The
result is
\begin{equation}
\sum_{|n| \leq l} |b^I_{d(ln)}(\nu_0)|^2 = \frac{2l+1}{4\pi}
I_0(l^2\sigma_d^2 e_d^2/2) {\rmn{e}}^{-l^2\sigma_d^2(1-e_d^2/2)},
\label{eq:52}
\end{equation}
so that
\begin{eqnarray}
\tilde{C}^{-1}_{(Ilm)(Il'm')} &=& \delta_{ll'}\delta_{mm'} \sum_d
\Bigl[ w_d I_0(l^2\sigma_d^2 e_d^2/2)\nonumber \\
&&\phantom{ \delta_{ll'}\delta_{mm'} \sum_d}
\times {\rmn{e}}^{-l^2\sigma_d^2(1-e_d^2/2)} \Bigr].
\label{eq:53}
\end{eqnarray}

If a known beam asymmetry is to be accounted for with pixel-based map-making
techniques, it is necessary to include the asymmetry in the real-space
pointing matrix. A simpler procedure suggests itself for the case of
randomly positioned rings, where we can exploit the fact that every pixel is
sampled with each detector in every orientation. Averaging the data
in each pixel from a given detector gives an unbiased estimate of the
pixelised sky smoothed with an effective window function. This window
function follows from azimuthally averaging the beam, so that
\begin{equation}
W_{{\rmn{eff}},l,d} \equiv \sqrt{\frac{4\pi}{2l+1}} b^{I}_{d(l0)}
= I_0(l^2 \sigma_d^2 e_d^2/4){\rmn{e}}^{-l^2\sigma_d^2(1-e_d^2/2)/2}.
\label{eq:54}
\end{equation}
The analysis of the smoothed maps proceeds as in
Section~\ref{subsec:random}, so that the inverse covariance matrix
evaluates to
\begin{eqnarray}
\tilde{C}^{-1}_{(Ilm)(Il'm')} &=& \delta_{ll'}\delta_{mm'} \sum_d
\Bigl[ w_d I_0^2(l^2\sigma_d^2 e_d^2/4)\nonumber \\
&&\phantom{ \delta_{ll'}\delta_{mm'} \sum_d}
\times {\rmn{e}}^{-l^2\sigma_d^2(1-e_d^2/2)} \Bigr].
\label{eq:55}
\end{eqnarray}
Clearly there is some information loss on averaging the data in the manner
described. For a single detector, the ratios of the diagonal elements of the
covariance matrices obtained from equations~(\ref{eq:53}) and (\ref{eq:55}) are
\begin{equation}
\frac{I_0^2(l^2\sigma_d^2e_d^2/4)}{I_0(l^2\sigma_d^2e_d^2/2)} =
1 - \frac{1}{32} (l\sigma_d e_d)^4 + \cdots . 
\end{equation}
This
information loss is only significant on scales below that of the beam asymmetry
($\approx \sigma_d e_d$). For more general scanning, averaging data in a pixel
will necessarily produce a biased estimate of the sky in the presence of a
beam asymmetry. This will be most acute for scanning strategies where a large
number of pixels are sampled with the detectors in only a narrow range of
orientations. Such strategies include constant latitude scanning, to which
we now turn.

\subsection{Constant latitude rings}
\label{subsec:latitude}

To analyse constant latitude scanning, we consider equation~(\ref{eq:35})
under the restriction $\theta_r=\theta$. Following the discussion
at the start of Section~\ref{subsec:random}, we consider the continuum
limit. For constant latitude scans we replace the sum over rings by
$[N_{{\rmn{r}}}/(2\pi)] \int \ud \phi_r$.
Expanding the $D$-matrices as in equation~(\ref{eq:9})
and performing the integral over $\phi_r$, the expression for
$\tilde{\mathbfss{C}}^{-1}$ reduces to
\[
\tilde{C}^{-1}_{(Plm)(P'l'm')} = 4\pi \delta_{mm'}
\sum_{dn} w_d d^l_{mn}(\theta) d^{l'}_{m'n}(\theta) \clb^{P}_{d(ln)}
\clb^{P'\ast}_{d(l'n)},
\]
\begin{equation}
\label{eq:56}
\end{equation}
where the sum is over detectors and $|n| \leq {\rmn{min}}(l,l')$. 
We obtain uncorrelated errors between modes with different $m$ due to
the azimuthal symmetry of the sky coverage~\cite{oh99}. Since
$\tilde{\mathbfss{C}}^{-1}$ is block-diagonal, it can be inverted
in $O(l_{{\rmn{max}}}^4)$ operations. In practice, variations in the
instrument parameters through the mission will spoil this exact block-diagonal
structure, as will any precession in the latitude $\theta$. However,
approximating $\tilde{\mathbfss{C}}^{-1}$ by its diagonal blocks may still
provide an adequate preconditioner for e.g.\ a conjugate gradient
reconstruction of the sky.

In equation~(\ref{eq:56}), $w_d \equiv T_{{\rmn{m}}}/(4\pi s_d^2)$ refers to
the weight per  solid angle for uniform coverage of the whole sky, which
differs from the pixel-dependent $w_{d,p}$ due to the variation in time spent
per solid angle. By differentiating the geometric relation
\begin{equation}
\cos\theta_p = \cos\alpha_d \cos\theta - \sin\alpha_d \sin\theta \cos\psi,
\label{eq:57}
\end{equation}
which relates the polar angle $\theta_p$ of the pixel containing the main beam
to the ring phase $\psi$, it is straightforward to show that
\begin{equation}
w_{d,p} = \frac{2}{\pi} \frac{\Theta(B_{d,p})}{\sqrt{B_{d,p}}} w_d,
\label{eq:58}
\end{equation}
where $\Theta(x)$ is the Heaviside unit step function, and we have
introduced the quantity
\begin{equation}
B_{d,p} \equiv 1- \cos^2\alpha_d - \cos^2\theta_p -\cos^2\theta + 2\cos\alpha_d
\cos\theta_p \cos\theta
\label{eq:59}
\end{equation}
for later convenience.

To proceed further, we make the assumption that the detector
sensitivities and ring opening angles are all equal, so that
$w_d = w$, $\alpha_d = \alpha$, and $B_{d,p}=B_p$. It follows that
the pixel-dependent weight per solid angle is also independent of detector:
$w_{d,p}=w_p$. The sum over detectors in equation~(\ref{eq:56}) then reduces to
\begin{eqnarray}
\sum_d \clb^{P\ast}_{d(ln)} \clb^{P'}_{d(l'n)} &=& \sum_{m''m'''}
[d^l_{nm''}(\alpha) d^{l'}_{n m'''}(\alpha) \nonumber \\
&& \mbox{} \times \sum_d
\tilde{b}^{P}_{d(lm'')} \tilde{b}^{P'\ast}_{d(l'm''')}]. 
\label{eq:60}
\end{eqnarray}
We further assume that all detectors have axisymmetric, co-polar
beams which are equivalent up to rotations about the beam axis. In this case,
the multipoles are given by equations~(\ref{eq:39})--(\ref{eq:41}) with the
angles $\rho_d$ determining the relative polarization orientations.
The choice of angles $\rho_d$ has a strong impact on the covariance structure
of the estimated multipoles. Designing the focal plane so that
$\sum_d \exp(2 i \rho_d) = 0$ ensures that the errors are uncorrelated between
the total intensity and the linear polarization.
The correlation between the $G$ and $C$
multipoles is controlled in part by the sum $\sum_d \exp(4 i \rho_d)$, although
demanding that this sum vanish is not a sufficient condition to
ensure uncorrelated errors between $G$ and $C$. However, the condition
$\sum_d \exp(4 i \rho_d)=0$ does ensure uncorrelated errors between
the smoothed Stokes fields\footnote{The smoothed fields are defined in
equations~(\ref{eq:44}) and (\ref{eq:45}). We have dropped the subscript
$d$ since we are assuming that the beam window functions are the same for all
detectors.} $Q_{{\rmn{eff}}}(\be)$ and $U_{{\rmn{eff}}}(\be)$
which, under the stringent restrictions adopted in this section, can be 
estimated in an optimal manner by a $\chi^2$ fitting of the data from all
detectors obtained when the main beam of each lies in the given
pixel~\cite{couchot99}. The smoothed maps estimated in this manner have
uncorrelated errors between pixels with weights per solid angle
$N_{{\rmn{d}}} w_{p}$ for $I_{{\rmn{eff}}}$, and
$N_{{\rmn{d}}} w_{p}/2$ for $Q_{{\rmn{eff}}}$ and $U_{{\rmn{eff}}}$, where,
recall, $N_{{\rmn{d}}}$ is the number of detectors. Following Couchot et
al.~\shortcite{couchot99}, we denote arrangements of polarimeters with
$\sum_d \exp(2 i \rho_d)=0$ and $\sum_d \exp(4i\rho_d)=0$ as optimal
configurations. In such a configuration, the non-vanishing components of the
inverse covariance matrix evaluate to
\begin{eqnarray}
\tilde{C}^{-1}_{(lmI)(l'm'I)} &=& \delta_{mm'} w N_{{\rmn{d}}} W_l W_{l'}
\sqrt{(2l+1)(2l'+1)} \nonumber \\
&&\times \sum_n d^l_{mn}(\theta) d^{l'}_{m'n}(\theta)
d^l_{n0}(\alpha) d^{l'}_{n0}(\alpha), \label{eq:61} \\
\tilde{C}^{-1}_{(lmG)(l'm'G)} &=& \delta_{mm'} w N_{{\rmn{d}}} \, {}_2W_l
\, {}_2W_{l'} \sqrt{(2l+1)(2l'+1)} \nonumber \\
&&\times \sum_n \{d^l_{mn}(\theta) d^{l'}_{m'n}(\theta)
{\textstyle{\frac{1}{2}}} [d^l_{n2}(\alpha) d^{l'}_{n2}(\alpha) \nonumber \\
&&\mbox{}\phantom{\sum_n \{}+ d^l_{n\, -2}(\alpha) d^{l'}_{n\, -2}(\alpha)]\},
\label{eq:62} \\
\tilde{C}^{-1}_{(lmG)(l'm'C)} &=& -i \delta_{mm'} w N_{{\rmn{d}}} \, {}_2W_l
\, {}_2W_{l'} \sqrt{(2l+1)(2l'+1)} \nonumber \\
&&\times \sum_n \{d^l_{mn}(\theta) d^{l'}_{m'n}(\theta)
{\textstyle{\frac{1}{2}}} [d^l_{n2}(\alpha) d^{l'}_{n2}(\alpha) \nonumber \\
&&\mbox{}\phantom{\sum_n \{}- d^l_{n\, -2}(\alpha) d^{l'}_{n\, -2}(\alpha)]\},
\label{eq:63}
\end{eqnarray}
with $\tilde{C}^{-1}_{(lmC)(l'm'C)}=\tilde{C}^{-1}_{(lmG)(l'm'G)}$.
In equations~(\ref{eq:61})--(\ref{eq:63}), the sums are over
$|n| \leq {\rmn{min}}(l,l')$.

We show in Appendix~\ref{app:map} that the sum over $n$ on the right-hand sides
of equations~(\ref{eq:61})--(\ref{eq:63}) can be reduced to matrix elements
of the time spent per solid angle with the appropriate spin weight basis:
\begin{eqnarray}
&& \delta_{mm'} \sum_{n} \sqrt{(2l+1)(2l'+1)}
d^l_{mn}(\theta) d^{l'}_{m'n}(\theta) d^l_{ns}(\alpha) d^{l'}_{ns}(\alpha)
\nonumber \\
&& \mbox{} = \int \frac{2}{\pi} \frac{\Theta(B_p)}{\sqrt{B_p}}
{}_{-s}Y_{(lm)}^\ast(\be_p)\, {}_{-s}Y_{(l'm')}(\be_p) \, \ud\Omega_p,
\label{eq:64}
\end{eqnarray}
for integer $s$. Using this result in equations~(\ref{eq:61})--(\ref{eq:63}),
and recalling equation~(\ref{eq:58}), the components of the inverse covariance
matrix can be reduced to
\begin{eqnarray}
\tilde{C}^{-1}_{(Ilm)(Il'm')} &=& W_{l}W_{l'} \int
w_{p} Y_{(lm)}^\ast Y_{(l'm')} \, \ud \Omega_p, \label{eq:65} \\
\tilde{C}^{-1}_{(Glm)(Gl'm')} &=& {}_2W_{l}\, {}_2W_{l'} \int[
w_{p} {\textstyle{\frac{1}{2}}}({}_2Y_{(lm)}^\ast {}_2Y_{(l'm')} \nonumber \\
&&\phantom{{}_2W_{l}\, {}_2W_{l'}}
+{}_{-2}Y_{(lm)}^\ast{}_{-2}Y_{(l'm')}) \, \ud\Omega_p] , \label{eq:66} \\
\tilde{C}^{-1}_{(Glm)(Cl'm')} &=& i\, {}_2W_{l}\, {}_2W_{l'} \int[
w_{p} {\textstyle{\frac{1}{2}}}({}_2Y_{(lm)}^\ast {}_2Y_{(l'm')} \nonumber \\
&&\phantom{{}_2W_{l}\, {}_2W_{l'}}
- {}_{-2}Y_{(lm)}^\ast{}_{-2}Y_{(l'm')}) \, \ud\Omega_p] .\label{eq:67}
\end{eqnarray}
Note that these results automatically take account of incomplete
sky coverage, through the presence of $\Theta(B_p)$. The function
$B_p < 0$ in pixels that are not sampled by the main beam during the
mission, so such pixels make no contribution to the covariance matrix.
As noted in Section~\ref{subsec:ml}, the presence of unsampled regions can
render $\tilde{\mathbfss{C}}^{-1}$ numerically singular. For the \emph{Planck}
mission the nominal ring opening angle $\alpha \approx 85\degr$, which would
leave $5\degr$ holes at the ecliptic poles for constant latitude scanning
in the ecliptic. However, a proposal to precess the spin axis out of
the ecliptic plane with an amplitude of $10\degr$ has the benefit of providing
complete sky coverage, which would render $\tilde{\mathbfss{C}}^{-1}$
invertible (although no longer block-diagonal).

To compare equations~(\ref{eq:65})--(\ref{eq:67}) with the results obtained
from a pixelised map, we can make use of
equations~(\ref{eq:47})--(\ref{eq:49}) which, it will be
recalled, give the errors on the maximum-likelihood solution for the
multipoles obtained from $N_{{\rmn{d}}}$ sets of smoothed maps. Here, we
have only one map (estimated using data from all detectors
as described above), with weights per pixel
$N_{{\rmn{d}}} w_p$ for $I_{{\rmn{eff}}}$ and $N_{{\rmn{d}}} w_p/2$ for
$Q_{{\rmn{eff}}}$ and $U_{{\rmn{eff}}}$. With these conditions, it is
straightforward to verify that
equations~(\ref{eq:47})--(\ref{eq:49}) reduce to pixelised versions
of equations~(\ref{eq:65})--(\ref{eq:67}). This observation confirms the
statistical equivalence of the map-based and harmonic routes through to the
multipoles of the sky for constant latitude scanning, under the strong
restrictions adopted in this section.

More generally, since the maps and their multipoles contain
the same information (disregarding pixelisation errors), it will always
be the case that the optimal map-making and map-to-multipole algorithms
are statistically equivalent to the optimal harmonic estimate of the
multipoles. However, as demonstrated in Section~\ref{sec:hdm}, the harmonic
model provides a more natural framework for the inclusion of a number of
systematic effects in the instrument modelling and data analysis
pipeline. The inclusion of some such effects is essential to maintain
the integrity of the final data products.

\section{Discussion}
\label{sec:disc}

In this section we discuss a number of important issues that arise during the
map-making stage. The treatment of low frequency noise (`destriping') in the
harmonic model is described in Section~\ref{subsec:stripes}. In
Section~\ref{subsec:sync} we comment on the control of
scan-synchronous instrument effects, focusing on straylight from bright
sources picked up through the sidelobes of the telescope, and the modulation
of the dipole in the rest-frame of the experiment. Finally, the overall
calibration of the experiment to external standards is described in
Section~\ref{subsec:gain}.

\subsection{Destriping}
\label{subsec:stripes}

In Section~\ref{subsec:noise} we established that for large $N_{{\rmn{s}}}$,
the main impact of noise power at frequencies below the spin frequency
is concentrated at the $n=0$ Fourier modes of the phase-ordered data (see
also, e.g.\ Delabrouille 1998). If the noise power varies sufficiently rapidly
in the range $0 \leq \omega \leq \omega_r/N_{{\rmn{s}}}$ these offsets may be
correlated between rings. Careful treatment of low frequency noise is
essential to avoid undesirable long range noise correlations
(`stripes') in the maps. Given the difficulty of inverting the time-time
noise covariance matrix in the presence of low frequency noise, additional
pre-processing steps are usually performed prior to map-making.
Pre-whitening the time- or phase-ordered data (e.g.\ Tegmark 1997) uses
a prior knowledge of the noise power spectrum, $\cln_d(\omega)$, to represent
the data in a basis where there are no noise correlations. In essence, this
involves the subtraction of an optimal offset from each ring of data.
Natoli et al.~\shortcite{natoli01} have successfully applied this method
to the 30 GHz channel of the \emph{Planck} Low Frequency Instrument under the
assumption of a symmetric beam profile. Other destriping methods advocated for
the \emph{Planck} mission, e.g.\ Delabrouille~\shortcite{del98a} and Revenu et
al.~\shortcite{revenu00}, also involve the subtraction of offsets from the
rings, but these offsets are obtained directly from the data by exploiting
the intersections of rings.

The harmonic model represents the phase-ordered data in the Fourier basis
on rings which ensures that the noise covariance matrix, $\mathbfss{N}$,
is very sparse, even in the presence of significant low frequency power.
This removes the need for any additional pre-whitening. A potential problem
arises if the noise power spectrum is not known accurately at low
frequencies, since this may compromise the quality of the sky
reconstruction. The harmonic model suggests a simple solution to this
problem: remove the $n=0$ Fourier modes from the analysis. From a Bayesian
viewpoint, we must now infer the $\bar{a}^P_{(lm)}$ from a knowledge
of $t_{(rdn)}$ with $n>0$. This inversion requires the probability density
function (pdf) for $n_{(rdn)}$ with $n>0$, which is obtained from the full
pdf by integrating over the $n=0$ modes. However, for Gaussian noise, the
pdf factors into products of pdfs for the individual Fourier modes as a
consequence of equation~(\ref{eq:24}), so that integrating over the $n=0$
modes is trivial. Note also that since the $n=0$ modes of the data are the
only ones that depend on the $l=0$ modes of the total intensity and circular
polarization, these monopole modes can no longer be reconstructed.
To implement the scheme one can either reformulate the problem with the
$n=0$ data modes removed at the outset, or, equivalently, let
$N_{(rd0)(r'd0)} \rightarrow \infty$, and perform the inversion in the
subspace orthogonal to $\bar{a}^I_{(00)}$ and $\bar{a}^V_{(00)}$.
Removing the $n=0$ modes
does not bias the solution, but it is no longer optimal given all the data,
since $t_{(rd0)}$ receives a contribution from all multipoles. However, for
large $l_{\rmn{max}}$ the information loss should be small for the
$l>0$ multipoles. Excluding the $n=0$ modes provides a very robust, simple way
of removing undesirable long range correlations in the maps that might arise
if the low frequency
noise were estimated inaccurately. Note also that this `destriping' method
handles the polarization signal seamlessly, avoiding the technicalities
involved in estimating offsets from ring intersections for polarized
detectors~\cite{revenu00}. Removing the $n=0$ and $l=0$ modes from the
analysis also neatly solves the
problem of degeneracy that arises if the experiment is insensitive to
the monopole -- this situation may arise for the \emph{Planck} HFI since the
favoured readout electronic system is insensitive to very low
frequencies~\cite{gaertner97}.

\subsection{Control of Scan-Synchronous Instrument Effects}
\label{subsec:sync}

Instrument effects that produce signals synchronous with the rotation of the
instrument will not be suppressed by co-adding the time-ordered data to form
the phase-ordered data. Given complete knowledge of the instrument
response, these potential
sources of systematic error can be controlled by adopting a more refined model
of the instrument response during the reconstruction of the sky. In this
subsection we discuss the removal of two potential systematic effects
within the context of the harmonic data model.

\subsubsection{Sidelobe Reconstruction}

An important source of systematic error is straylight from
e.g.\ the Galaxy and the CMB dipole entering the
instrument through the sidelobes of the telescope. Here,
the difficulty lies not in the inclusion of the sidelobes in the reconstruction
process, since the harmonic model allows for a complete description of the
beams, but rather in the limited knowledge of the sidelobes that will be
available from simulations and from in-flight
calibrations~\cite{burigana01,van01}. Although the
sidelobes can be expected to be highly polarized, the direction should be
sufficiently random that it is adequate to model only the directivity
$\tilde{I}_d(\be;\nu_0)$ in the sidelobes.

Delabrouille et al.~\shortcite{del98c} have proposed an iterative scheme for
estimating sidelobe corrections. In their method, an estimate of the sky is
obtained from a first pass of the map-making algorithm ignoring sidelobe
corrections, and this is then used
to compute the difference between the observed time-ordered data and the
contribution of the sky coming through the main beam. This difference is
attributable to instrument noise and the sidelobe signal, and can be inverted
(with suitable regularisation) to estimate the values of the directivity
in the sidelobe pixels. With this improved knowledge of the beam, and the
original estimate of the sky, an estimate of the sidelobe contribution can
be subtracted from the time-ordered data, and an improved estimate of the sky
obtained from a further pass of the original map-making algorithm (again with
sidelobes ignored). This process can be iterated to give consistent
estimates of the sky and the directivity in the sidelobe. It should be
possible to implement a similar scheme in the harmonic model, with the
sidelobes parameterised by a modest number of multipoles
and the sidelobe contribution estimated from the $t_{(rdn)}$, although we have
not attempted this yet. One potential problem with such a scheme is that the
sidelobes may contain rather sharp features due to geometrical effects, in
which case a parameterisation of the sidelobes with spherical multipoles may
not be ideal. The monopole $b^I_{(00)}(\nu_0)$ should not be varied during
sidelobe reconstruction since its value is fixed at $1/\sqrt{4\pi}$ by
definition.
The iterative reconstruction of the sidelobes does not require an absolute
calibration of the detector gains, and should be performed prior to
the calibration procedures described in Section~\ref{subsec:gain}.

\subsubsection{Dipole variation}

One subtlety that we have overlooked so far, which is significant for
survey missions from space, is the variation of satellites (linear) velocity
during the year. The orbital velocity of the earth about the sun $\sim 30\,
{\rmn{kms}}^{-1}$, which induces a yearly modulation in the multipoles seen in
the satellite's rest frame. For sensitivities equivalent to the \emph{Planck}
mission (a few $\mu {\rmn{K}}/K$ in $100\, {\rmn{arcmin}}^2$ pixels at
100 GHz, for twelve months of observation), the most significant
effect is the modulation of the intensity dipole due to the monopole
with amplitude $\sim 10^{-4}\, {\rmn{K}}$. If we denote the multipoles
in the frame of the satellite by $a^P_{(lm),{\rmn{E}}}(\nu)$,
and continue to denote the multipoles on the
$\{\bsigma_x,\bsigma_y,\bsigma_z\}$ basis fixed relative to the solar system
by $a^P_{(lm)}(\nu)$, we have
\begin{eqnarray}
a^P_{(lm),{\rmn{E}}}(\nu) &\approx& a^P_{(lm)}(\nu) + \delta^P_I \delta_{l1}
\frac{\nu^4}{\sqrt{4\pi}} \frac{\ud}{\ud\nu}\left[\frac{a^I_{(00)}(\nu)}{\nu^3}
\right] \nonumber \\
&&\mbox{}\times\int {\bmath{v}}\cdot \bmath{e}
Y_{(lm)}^\ast(\bmath{e}) \, \ud \Omega,
\label{eq:68}
\end{eqnarray}
where $\bmath{v}$ is the (time-dependent) velocity of the earth relative to
the sun, and $\bmath{e}$ is a unit vector. The $a^P_{(lm),{\rmn{E}}}(\nu)$
are defined relative to a basis obtained by Lorentz boosting the
$\{\bsigma_x,\bsigma_y,\bsigma_z\}$ basis. Aberration effects on the
high $l$ multipoles may also be significant for
\emph{Planck} (Challinor \& van Leeuwen, in preparation).

It is the time-dependent multipoles $\bar{a}^P_{(lm),{\rmn{E}}}$ (obtained by
averaging $a^P_{(lm),{\rmn{E}}}(\nu)$ across the spectral filter)
which should now appear in equation~(\ref{eq:20}), relating the
signal to the sky. However, we must still solve for the time-independent
multipoles $\bar{a}^P_{(lm)}$. If the monopole $a^I_{(00)}(\nu)$ were available
at the start of the analysis for all frequency channels, the contribution
of the time-dependent part of $a^P_{(lm),{\rmn{E}}}(\nu)$ to the signal could
be subtracted from the $t_{(rdn)}$ prior to solving for the multipoles.
Note that this process only affects the $n=0$ and $n=1$ Fourier modes,
$t_{(rdn)}$, but these influence all multipoles of the reconstructed sky.
More realistically, the monopole $a^I_{(00)}(\nu)$ may only be poorly
determined prior to the analysis, in which case it will be necessary
to iterate the estimation of the multipoles, using the monopole determined
in the previous iteration to correct the $t_{(rdn)}$. This process is clearly
not possible if the $n=0$ modes are ignored in the analysis, as may be
necessary for the reasons discussed
in Section~\ref{subsec:stripes}. A radical way to circumvent the problems of
poorly determined low frequency noise and dipole modulation is to remove the
$n=1$ modes from the analysis also. However, an internally reconstructed dipole
appears to be essential for establishing an absolute normalisation of the
experiment (see next subsection).

\subsection{Gain calibration}
\label{subsec:gain}

The methods described by van Leeuwen et al.~\shortcite{van01} for producing
internally consistent ring-sets do not provide an absolute gain calibration
for each detector. By forgoing the introduction of any external flux
calibrator during the reconstruction of the ring-sets, only the ratios of the
gains of all detectors in a given frequency band will be known;
the absolute gains will only be determined up to an overall factor,
$g$. This factor is defined to be the ratio of the true gains,
$G_{d,{\rmn{true}}}$, to those assumed in the model of the instrument, $G_d$
(e.g.\ equation~\ref{eq:19}). An external calibrator provides prior information
on the sky, ${\rmn{Pr}}(\bmath{a})$, which can be used to constrain $g$. For
simplicity, we assume a Gaussian prior:
\begin{equation}
{\rmn{Pr}}(\bmath{a}) \propto | \pi {\mathbfss{C}}_{\rmn{ext}} |^{-1/2}
\exp[-(\bmath{a}-\bmath{a}_{\rmn{ext}})^\dagger {\mathbfss{C}}_{\rmn{ext}}^{-1}
(\bmath{a}-\bmath{a}_{\rmn{ext}})/2],
\label{eq:69}
\end{equation}
where $\bmath{a}_{\rmn{ext}}$ is the most likely prior sky, and
$\mathbfss{C}_{\rmn{ext}}$ is the associated error. The sky and normalisation
can be now estimated directly from the (Fourier) ring data, $\bmath{t}$,
by maximising ${\rmn{Pr}}(\bmath{t}|\bmath{a},g){\rmn{Pr}}(\bmath{a})$
(assuming a uniform prior on $g$). In the usual limit where the
experiment determines the sky to much better precision than the calibrator,
the best estimate of the sky is $\hat{\bmath{a}}/\hat{g}$, where
$\hat{\bmath{a}}$ is the maximum-likelihood estimate for the sky given in
equation~(\ref{eq:28}), obtained with gains $G_d$, and the best estimate
of the normalisation is given by
\begin{equation}
\hat{g} = \frac{\hat{\bmath{a}}^\dagger {\mathbfss{C}}_{\rmn{ext}}^{-1}
\hat{\bmath{a}}}{\Re(\hat{\bmath{a}}^\dagger {\mathbfss{C}}_{\rmn{ext}}^{-1}
\bmath{a}_{\rmn{ext}})}.
\label{eq:70}
\end{equation}
In practice, ${\mathbfss{C}}_{\rmn{ext}}^{-1} \bmath{a}_{\rmn{ext}}$
and ${\mathbfss{C}}_{\rmn{ext}}^{-1}$ may have to be estimated from
existing observations on patches of the sky. In such cases,
equation~(\ref{eq:70}) reduces to the solution obtained by
minimising the (error-weighted) residuals between the reconstructed map,
$\sum_{lm} g^{-1} \hat{a}^I_{(lm)} Y_{(lm)}(\bmath{e})$, and the calibration
map. The requirement that the calibration maps be at the same frequencies
as the instrument channels can be removed by selecting a large patch, at
high Galactic latitude, where the large-scale signal is dominated by the CMB
dipole.

\section{Further analysis of frequency maps}
\label{sec:sub}

Although the main focus of this paper is the reconstruction of frequency maps
of the sky in multipole space, in this section we offer some comments on the
subsequent analysis of these maps into their astrophysical components, and
estimated power spectra.

\subsection{Component separation}
\label{subsec:component}

The frequency maps contain contributions from a number of astrophysical
components which must be separated on the basis of their assumed frequency
spectra (and possibly power spectra).
Unresolved radio sources require separate processing since they cannot
be accurately modelled as a population with a single frequency spectrum.
If bright point sources were not removed from the ring-sets prior to
map-making, their contribution
should be filtered from the multipoles before attempting component
separation~\cite{vielva01}. Within the timescale of \emph{Planck}, the majority
of point sources expected to be visible in the 857
GHz channel of the HFI should already have been catalogued prior to launch
by the \emph{ASTRO-F} survey~\footnote{http://www.ir.isas.ac.jp/ASTRO-F/}. Such
catalogues will be valuable for the geometrical calibration of the
ring-sets using the positions of identified point-sources~\cite{van01}, and
also for the removal of point sources from the reconstructed frequency maps.
If bright point sources have already been removed from the ring-sets prior to
map-making, component separation can proceed directly from the reconstructed
multipoles. However, statistical
reconstruction of faint point sources may still be desirable, in which case
the faint point sources can be included as a generalised noise
in the separation algorithm for the other astrophysical
components~\cite{hobson99}.

The full-sky, maximum-entropy component separation algorithm developed
recently by Stolyarov et al.~\shortcite{stol01} performs the separation in the
spherical multipole basis, so that it can use the outputs of the multipole
estimation process described here directly. Although the
algorithm of Hobson et al.\ does not handle polarization in its current
form, the extension to polarized components should be straightforward.
(The Wiener separation of polarized components has been implemented
successfully by Bouchet, Prunet \& Sethi 1999 for small patches of sky.) A more
demanding problem for high resolution, all-sky separation algorithms
is the inclusion of non-isotropic noise, since then the separation cannot
be performed multipole by multipole due to the correlated errors
(e.g.\ Prunet et al.\ 2001).

\subsection{Power spectrum estimation}
\label{subsec:power}

Power spectrum estimation can also be performed efficiently in multipole
space. The simplest, unbiased estimator which is quadratic in the
estimated dimensionless multipoles $\tilde{a}^P_{(lm)}$ is of the form
\begin{equation}
\hat{C}_l^{PP'} = \frac{1}{(2l+1)}\sum_m(\tilde{a}^P_{(lm)}
\tilde{a}^{P'\ast}_{(lm)}
- \tilde{C}_{(Plm)(P'lm)}).
\label{eq:71}
\end{equation}
The variance of this estimator is calculated in Kamionkowski et
al.~\shortcite{kamion97} under the
assumptions of full sky coverage and isotropic pixel noise which is
uncorrelated between Stokes parameters. In this case, the quadratic estimator
in equation~(\ref{eq:71}) is equivalent to the maximum-likelihood estimator.

For more general noise properties and scan strategies the estimator
in equation~(\ref{eq:71}) is sub-optimal. Maximum-likelihood techniques have
been developed which allow accurate computation of the temperature power
spectrum in $O(l_{{\rmn{max}}}^4)$ operations for scanning strategies with
(near) azimuthal symmetry~\cite{oh99}. This method cannot be directly in the
context of the harmonic model
since Oh et al.\ store the inverse (noise) covariance matrix,
${\mathbfss{C}}^{-1}$, in the pixel basis only,
where they assume it is diagonal. Forward and inverse fast
spherical transforms are then employed to apply ${\mathbfss{C}}^{-1}$
efficiently in multipole space. Applying ${\mathbfss{C}}^{-1}$ directly
in multipole space would not increase the overall operations count
significantly, but would increase the memory requirements to
$O(l_{{\rmn{max}}}^4)$ in general.

One potential issue in power spectrum estimation is the treatment of regions
near to the Galactic plane. Although component separation algorithms
can reconstruct the CMB very well even at low Galactic
latitude (Stolyarov et al.\ 2001), it may still be necessary to
remove traces of the Galaxy by brute force. Given a pixelised map, the pixels
near the plane can be excised from the analysis before estimating the
multipoles and their error covariance, and subsequent power spectrum
estimation. (Alternatively, power spectrum estimation can be performed directly
in real space away from the Galactic plane, although the lack of a sparse
signal covariance matrix in the pixel representation is problematic for high
resolution data-sets.)

Within the context of the harmonic model, the
Galactic plane can be dealt with by projecting the reconstructed
multipoles, $\hat{a}^P_{(lm)}$, into a subspace which is (almost) free from
Galactic contamination. For simplicity, we shall only describe the procedure
for a total power measurement; the extension to polarized data is
straightforward. We define the Hermitian matrix ${\mathbfss{P}}$ to have
components
\begin{equation}
P_{(lm)(l'm')} \equiv \int_{S^{2\prime}} Y_{(lm)}^\ast(\be)Y_{(l'm')}(\be)
\, \ud \Omega,
\label{eq:72}
\end{equation}
where the integral is over the region of the sky that we wish to retain
in the analysis. In the limit $\l_{\rmn{max}} \rightarrow \infty$,
$P_{(lm)(l'm')}$ are the matrix elements of a projection operator
(i.e.\ ${\mathbfss{P}}^2 = {\mathbfss{P}}$) which projects functions into
the region $S^{2\prime}$. For finite $\l_{\rmn{max}} \gg 1$, ${\mathbfss{P}}$
is almost a projection operator since the distribution of its eigenvalues
is almost bimodal with values clustered close to zero and one~\cite{mort01}.
The fraction of eigenvalues close to unity is approximately the fraction of the
sky retained; the remainder are nearly all zero. Performing
(maximum-likelihood) power spectrum estimation with the data object
${\mathbfss{P}} \hat{\bmath{a}}$ would not remove the Galactic contamination
since ${\mathbfss{P}}$ is (formally) invertible. To enforce rejection of the
Galaxy, we introduce a projection operator
$\tilde{{\mathbfss{P}}}$ which is obtained from ${\mathbfss{P}}$ by
retaining the eigenvectors but setting those eigenvalues greater
then $1-\epsilon$ to unity, and to zero otherwise.
The choice of threshold, $\epsilon\ll 1$, must be determined from
simulations to ensure good rejection of the Galaxy while minimising the
number of modes lost from the analysis. Working with
$\tilde{{\mathbfss{P}}}\hat{\bmath{a}}$ ensures that we have projected out
those modes of $\hat{\bmath{a}}$ which are (nearly) localised in the region
external to $S^{2\prime}$. This procedure is similar to power spectrum
estimation from the cut map, since projecting pixelised data into
$S^{2\prime}$ in real space has the effect of greatly amplifying the noise
on the multipoles estimated from the cut map along those directions in
multipole space which correspond to functions nearly localised in the cut.
In practice, it is likely to be more efficient to represent the projected
data vector $\tilde{{\mathbfss{P}}}\hat{\bmath{a}}$ on a basis adapted to the
region $S^{2\prime}$, so that we work with the object ${\mathbfss{U}}^\dagger
\hat{\bmath{a}}$, where ${\mathbfss{U}}$ is the (non-square) matrix whose
columns are the eigenvectors of $\tilde{{\mathbfss{P}}}$ with unit eigenvalue.
The matrix ${\mathbfss{U}}$ can easily be obtained from a singular value
decomposition of $\tilde{{\mathbfss{P}}}$~\cite{mort01}.

\section{Conclusion}
\label{sec:conc}

Most current techniques for analysing CMB data on the sphere are based on
the use of pixelised maps. The basis for a complementary approach,
based on spherical harmonic coefficients of the intensity and polarization,
has been
derived here for the case of experiments that scan the full sky in circles.
Our method offers a number of advantages, most notably the
refined treatment of non-ideal beam effects, the ability to handle correlated
(low frequency) noise in an optimal, but efficient, manner, and the
seamless way in which polarized data can be analysed alongside total
intensity data. In principle, harmonic methods can be used to develop
an end-to-end pipeline for the analysis of full-sky survey data,
allowing the clean
propagation of noise and other errors from the time-ordered data to the
CMB power spectra. In practice, the methods described here are best regarded
as being complementary to more standard map-based techniques rather than
a replacement. Undoubtly, there are analysis projects that are better
suited to map-based techniques -- typically those concerned with the science
of local features in the maps, such as foregrounds. In addition, power
spectrum estimation becomes cumbersome with purely harmonic methods if we
have reason to question the foreground contamination of certain linear
combinations of the spherical multipoles (e.g.\ those corresponding to
features localised in the galactic plane).

The biggest hurdle facing a practical implementation of the harmonic
`map'-making method at the resolution demanded by the upcoming satellite
missions, is the need to solve the $O(l_{{\rmn{max}}}^2)\times
O(l_{{\rmn{max}}}^2)$ linear system
in equation~(\ref{eq:28}). Unlike conventional map-making in the presence of
noise correlations, the problem lies not in the inversion of the
noise covariance matrix, since we are working in a representation where
this matrix is already very sparse. The inversion of
${\mathbfss{A}}^\dagger {\mathbfss{N}}^{-1} {\mathbfss{A}}$ can be avoided by
adopting iterative techniques with a block-diagonal preconditioner.
The main computational overhead arises from computing and storing the
large, non-sparse matrix $\mathbfss{A}$ and applying it to the sky multipoles
and its transpose to the Fourier ring data. These
operations can be performed very efficiently using fast Fourier transform
techniques for the rather idealised case of constant latitude scanning,
with rings uniformally spaced in azimuth and no variations in instrument
properties during the mission~\cite{wan01b}. Unfortunately, it does not appear
that such techniques can be easily extended to more realistic scan
strategies. An assessment of some of these numerical problems, together with
a number of potential solutions, will be given in Mortlock et
al.\ (in preparation).

\section*{Acknowledgments}

This work benefited from useful discussions with several members of the
Cambridge Planck Analysis Centre, in particular Martin Bucher, Rob Crittenden,
and Neil Turok. ADC acknowledges a PPARC Postdoctoral Fellowship;
DJM and MAJA are funded by PPARC.

\appendix
\section{Main beam variation across a spectral band}
\label{app:beam}

In this appendix we discuss the systematic error that arises from ignoring
the variation of the beam multipoles with frequency across a spectral band
when analysing microwave data. This question does not appear to have received
much attention so far in the CMB literature.
For simplicity, we consider a single
detector which is only sensitive to the total intensity. We take the beam
profile to be an axisymmetric Gaussian, with frequency-dependent beam width
$\sigma(\nu)$, so that for $\sigma(\nu) \ll 1$ we have
\begin{equation}
b^I_{(lm)}(\nu) = \delta_{m0} \sqrt{\frac{2l+1}{4\pi}} W_l(\nu),
\label{eq:a1}
\end{equation}
where the scalar window function is
\begin{equation}
W_l(\nu)=\exp[-l(l+1)\sigma^2(\nu)/2].
\label{eq:a2}
\end{equation}
Assuming that the beam is diffraction limited, we expect the frequency
dependence of the beam width to go like $\sigma(\nu) = (\nu_0/\nu)
\sigma_0$, where $\sigma_0 \equiv \sigma(\nu_0)$ is the beam width
at the central frequency of the band. For $l \gg 1$ this gives
\begin{equation}
b^I_{(lm)}(\nu) = \sqrt{\frac{\nu}{\nu_0}} b^I_{(l\nu_0/\nu,m)}(\nu_0),
\label{eq:a3}
\end{equation}
which also holds more generally for arbitrary beams which scale inversely with
frequency along longitudes.

If the CMB is the dominant physical component in the frequency band
under consideration, the frequency spectrum of the brightness anisotropies
in linear theory is
\begin{equation}
a^I_{(lm)}(\nu) = \tilde{a}^I_{(lm)} \bar{T} \left. 
\frac{\partial B(\nu,T)}{\partial T} \right|_{T=\bar{T}},
\label{eq:a4}
\end{equation}
for $l>0$, where $B(\nu,T)$ is the Planck function and $\bar{T}$ is the
average CMB temperature over the sky. The $\tilde{a}^I_{(lm)}$, which describe
the dimensionless anisotropy in the thermodynamic temperature of the sky,
are independent of frequency for the CMB. We further restrict attention
to white noise, assume uniform coverage of the spin axis pointing over the
entire sky, and ignore any systematic variation in focal plane geometry.
Then, if we attempt to solve for the $\tilde{a}^I_{(lm)}$, but ignore the
variation of $W_l(\nu)$ with frequency, the maximum-likelihood solution
returns a biased estimate $\hat{a}^I_{(lm)}$ which has the form
\begin{equation}
\hat{a}^I_{(lm)} = (1+\delta_l) \tilde{a}^I_{(lm)} + n_{(lm)},
\label{eq:a5}
\end{equation}
where the fractional bias
\begin{equation}
\delta_l \equiv \frac{\int W_l(\nu) (\partial B/\partial T)
v(\nu) \, \ud \nu}{W_l(\nu_0) \int (\partial B/\partial T)
v(\nu) \, \ud \nu} -1, 
\label{eq:a6}
\end{equation}
and the dimensionless random errors $n_{(lm)}$ have zero mean, and covariance
\begin{equation}
\langle n_{(lm)} n^\ast_{(l'm')} \rangle = \delta_{ll'}\delta_{mm'}
w^{-1} W_l^{-2}(\nu_0).
\label{eq:a7}
\end{equation}
Here, $w$ is the dimensionless weight per solid angle~(Knox 1995; see also
Section~\ref{sec:scan}). We can use the $\hat{a}^I_{(lm)}$ to estimate the
CMB power spectrum $C_l$ with the estimator
\begin{equation}
\hat{C}_l = \frac{1}{2l+1}\sum_{|m| \leq l} |\hat{a}^I_{(lm)} |^2
- w^{-1} W_l^{-2}(\nu_0),
\label{eq:a8}
\end{equation}
which would be unbiased if $\delta_l$ were zero. The fractional
systematic error in our estimate of $C_l$ is therefore
\begin{equation}
(\Delta C_l/C_l)_{\rmn{syst}} = \delta_l(\delta_l+2), 
\label{eq:a9}
\end{equation}
while there is a fractional random error
\begin{equation}
(\Delta C_l/C_l)_{\rmn{rand}}= \sqrt{\frac{2}{2l+1}}[1+C_l^{-1} w^{-1} W_l^{-2}
(\nu_0)].
\label{eq:a10}
\end{equation}
The first term in brackets on the right-hand side of equation~(\ref{eq:a10})
is cosmic variance. If instead we asked how well we could reconstruct the
rotational-invariant $\sum_m|\tilde{a}^I_{(lm)}|^2/(2l+1)$ for our given
realisation of the sky, the cosmic variance term would not be present.
The latter situation is the more relevant for mapping other astrophysical
foreground components.

\begin{figure}
\epsfig{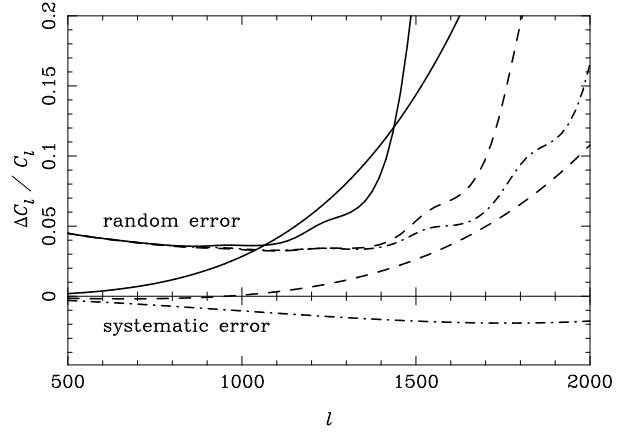}
\caption{The systematic error in the recovered temperature power spectrum
from neglecting the variation of the beam size across the spectral band
for the 100 (solid lines), 143 (dashed lines), and 217 GHz (dash-dotted)
\emph{Planck} HFI channels. Also plotted is the random error in the
$C_l$ from the combined instrument (white) noise from all detectors at the
specified frequency, and from cosmic variance. We have assumed that all beams
are axisymmetric with Gaussian profiles, and that the beam size is
diffraction limited.}
\label{fig:beam}
\end{figure}

In Fig.~\ref{fig:beam} we plot the systematic and random errors
on the CMB power spectrum for the 100, 143, and 217 GHz channels of the
\emph{Planck} HFI, using the predicted instrument specifications.
The spectral filter $v(\nu)$ is assumed to be a top hat with fractional
width $\Delta \nu/\nu_0 = 0.33$. It is clear that in the 100 GHz channel,
the systematic error due to our neglect of the variation of beam width
across the spectral band is non-negligible compared to the random error
due to instrument noise over a  broad range of $l$. For an axisymmetric beam
the effect can be easily accounted for by including the frequency-dependent
window function $W_l(\nu)$ in the integral over frequency, since the
integral $\int a^{P\ast}_{(lm)}(\nu) \tilde{b}^P_{d(lm')}(\nu) v(\nu)
\, \ud\nu$ would still factor into a part depending on the sky with
indices $l$ and $m$ only, as required for subsequent analysis. However,
for a non-axisymmetric beam where the variation of $\tilde{b}^P_{d(lm')}(\nu)$
cannot be reduced to a frequency-dependent factor with only an $l$ index, and
a constant part with $l$ and $m'$ indices, we can no longer integrate the
variation of beam with frequency and still preserve the factorisation of the
sky. Note that if there is only an effective beam asymmetry, arising from
skewing a symmetric beam with the temporal response of the instrument, the
above comments do not apply since the frequency dependence of the
effective beam will factor.

\section{Connecting ring-sets with maps}
\label{app:map}

In this appendix we establish the result given as equation~(\ref{eq:64})
in the main text, which relates the geometry of the ring-set to the
observing time per solid angle on the sky.

The first step is to replace the single summation $\sum_{|n| \leq {\rmn{min}}
(l,l')}$ on the left-hand side of equation~(\ref{eq:64}) with the double
summation $\sum_{l'' \geq {\rmn{max}}(|m|, |s|)} \delta_{l'l''}
\sum_{|n| \leq {\rmn{min}}(l,l'')}$, and to replace the labels $l'$ by
$l''$ in the argument of the summation. Denoting the left-hand side
of equation~(\ref{eq:64}) by ${}_sI^{ll'}_{mm'}$, we now have
\begin{eqnarray}
{}_sI^{ll'}_{mm'} &=& \sum_{l'' n} [\delta_{l'l''} \delta_{mm'}
\sqrt{(2l+1)(2l''+1)} \nonumber \\
&&\mbox{} \times d^l_{mn}(\theta) d^{l''}_{mn}(\theta)
d^l_{ns}(\alpha) d^{l''}_{ns}(\alpha)],
\label{eq:b1}
\end{eqnarray}
where we  have suppressed the limits on the summations. We now use the
orthonormality of the spin-weighted harmonics,
\begin{equation}
\int {}_{-s}Y_{(l'm')}(\be_p)\, {}_{-s}Y_{(l''m)}^\ast(\be_p)\, \ud\Omega_p
= \delta_{mm'} \delta_{l'l''},
\label{eq:b2}
\end{equation}
to replace the Kronecker deltas in equation~(\ref{eq:b1}) with the integral
of spin-weight $-s$ harmonics over the sphere. The term
${}_{-s}Y_{(l''m)}^\ast(\be_p)$ can be expressed in terms of $D$-matrices
using equation~(\ref{eq:46}):
\begin{equation}
{}_{-s}Y_{(l''m)}^\ast(\be_p) = (-1)^m \sqrt{\frac{2l''+1}{4\pi}}
D^{l''}_{sm}(0,\theta_p,\phi_p),
\label{eq:b3}
\end{equation}
where we have also used $d^l_{mm'}(\beta)=(-1)^{m-m'}d^l_{m'm}(\beta)$
(e.g.\ Brink \& Satchler 1993), so that
\begin{eqnarray}
{}_sI^{ll'}_{mm'} &=& (-1)^m \sqrt{\frac{2l+1}{4\pi}}
\int \sum_{l'' n}\{ {\rmn{e}}^{-im\phi_p}
\,{}_{-s}Y_{(l'm')}(\be_p) \nonumber\\
&&\mbox{}\times [(2l''+1) d^{l''}_{mn}(\theta)d^{l''}_{ns}(\alpha)
d^{l''}_{sm}(\theta_p)] \nonumber \\
&&\mbox{}\times d^l_{mn}(\theta)d^l_{ns}(\alpha)\}\, \ud \Omega_p .
\label{eq:b4}
\end{eqnarray}
If we now reverse the order of summation in equation~(\ref{eq:b4}) using
\begin{equation}
\sum_{l'' \geq {\rmn{max}}(|m|, |s|)} \,
\sum_{|n| \leq {\rmn{min}}(l,l'')} = \sum_{|n| \leq l} \,
\sum_{l'' \geq {\rmn{max}}(|m|,|n|,|s|)},
\label{eq:b5}
\end{equation}
the summation over $l''$ can be performed with the Ponzano-Regge sum rule
(e.g.\ Varshalovich, Moskalev \& Khersonskii 1988; equation 21, p.\ 89):
\begin{eqnarray}
&&\sum_{l'' \geq {\rmn{max}}(|m|,|n|,|s|)} (2l''+1)
d^{l''}_{mn}(\theta) d^{l''}_{ns}(\alpha) d^{l''}_{sm}(\theta_p) \nonumber \\
&&\phantom{xxxxx}
= \frac{2}{\pi}\frac{\Theta(B_p)}{\sqrt{B_p}}
\cos(m \delta_1 + n \delta_2 + s \delta_3),
\label{eq:b6}
\end{eqnarray}
where $B_p$ is given by equation~(\ref{eq:59}), 
and the angles $\delta_{1}$, $\delta_2$, and $\delta_3$ are defined
by the ${\rmn{SO}}(3)$ composition
\begin{equation}
D(\delta_1,\theta,0)D(\delta_2,\alpha,\delta_3)=D(0,-\theta_p,0).
\label{eq:b7}
\end{equation}
(The angles are given explicitly by Varshalovich et al.\ 1988, but it
is straightforward to show that their definitions are equivalent to the
implicit definition given here.) Finally, we can perform the remaining
sum over $n$ in equation~(\ref{eq:b4}) by writing $\cos(m\delta_1
+ n\delta_2 + s\delta_3)$ as the real part of $\exp[-i(m\delta_1
+n\delta_2 + s\delta_3)]$, and combining the complex exponentials with the
remaining $d$-functions to get
\begin{eqnarray}
&&\sum_{|n| \leq l} d^l_{mn}(\theta) d^l_{ns}(\alpha) \cos(m\delta_1
+ n\delta_2 + s\delta_3) \nonumber \\
&&\mbox{} = \Re \sum_{|n|\leq l}
D^l_{mn}(\delta_1,\theta,0)D^l_{ns}(\delta_2,\alpha,\delta_3) \nonumber \\
&&\mbox{} = d^l_{sm}(\theta_p,),
\label{eq:b8}
\end{eqnarray}
where we have used equation~(\ref{eq:b7}) in the last equality. The remaining
terms in equation~(\ref{eq:b4}) combine with $d^l_{sm}(\theta_p,)$ to give
${}_{-s}Y_{(lm)}^\ast(\be_p)$ on using equation~(\ref{eq:b3}). Our final
result is
\begin{equation}
{}_sI^{ll'}_{mm'} =  \int \frac{2}{\pi} \frac{\Theta(B_p)}{\sqrt{B_p}}
{}_{-s}Y_{(lm)}^\ast(\be_p)\, {}_{-s}Y_{(l'm')}(\be_p) \, \ud\Omega_p,
\label{eq:b9}
\end{equation}
which establishes equation~(\ref{eq:64}).

\bsp  
\label{lastpage}
\end{document}